\documentclass[english,11pt,aps,prd,a4paper,preprintnumbers, floatfix,nofootinbib,showpacs,superscriptaddress,  notitlepage]{revtex4-1} 
 \pdfoutput=1
\usepackage[usenames,dvipsnames]{color}  
\usepackage{graphicx}
\usepackage{setspace}
\usepackage{upgreek}
\usepackage{braket}
\usepackage{caption}
\usepackage{stackengine}
\usepackage{subcaption}
\usepackage{amsmath}
\usepackage{amssymb}
\usepackage[colorlinks=true,citecolor=darkred,urlcolor=darkred, pdfborder={0 0 0}]{hyperref}
\usepackage[normalem]{ulem}
\usepackage{xcolor}
\usepackage{graphicx}
\makeatletter
\def\p@subsection{}
\makeatother
\usepackage{xcolor}
\usepackage{float}

\usepackage{soul}

\definecolor{darkred}{rgb}{0.6,0,0}

\definecolor{linkcolor}{rgb}{0,0,0.5}

\definecolor{vdrgreen}{rgb}{0.0, 0.7, 0.0}

\newcommand{\stkout}[1]{\ifmmode\text{\sout{\ensuremath{#1}}}\else\sout{#1}\fi}

\def\gsim{\raise0.3ex\hbox{$\;>$\kern-0.75em\raise-1.1ex\hbox{$\sim\;$}}}
\def\lsim{\raise0.3ex\hbox{$\;<$\kern-0.75em\raise-1.1ex\hbox{$\sim\;$}}}

\def\beqn#1{\begin{equation}\label{#1}}
\def\eeqn{\end{equation}}

\def\beqa#1{\begin{eqnarray}\label{#1}}
\def\eeqa{\end{eqnarray}}

\usepackage{caption}
\usepackage{subcaption}
\usepackage{adjustbox}

\def\Z2{$\mathcal{Z_2}$}

\usepackage{bbold}
\usepackage{ragged2e}

\newcommand {\ignore}[1]{}

\def\cevns{CE$\nu$NS }
\def\eves{E$\nu$ES }
\usepackage{mathrsfs}

\makeatletter
\newcommand\footnoteref[1]{\protected@xdef\@thefnmark{\ref{#1}}\@footnotemark}
\makeatother

\def\321{$\mathrm{SU(3) \otimes SU(2) \otimes U(1)}$ }

\newcommand{\AddrAthens}{%
Department of Physics, National and Kapodistrian University
of Athens, Zografou Campus GR-15772 Athens, Greece}

\usepackage{appendix}
\begin{document}


\title{Probing generalized neutrino interactions with the DUNE Near Detector}

\author{P. Melas}\email{pmelas@fnal.gov}
\affiliation{\AddrAthens}

\author{D. K. Papoulias}\email{dkpapoulias@phys.uoa.gr}
\affiliation{\AddrAthens}

\author{N. Saoulidou}\email{Niki.Saoulidou@cern.ch}
\affiliation{\AddrAthens}

\begin{abstract}

We explore the prospects of constraining general non standard interactions involving light mediators through elastic neutrino-electron scattering events at the DUNE Near Detector (ND). We furthermore consider the special cases of light vector mediators in motivated models such as $U(1)_{B-L}$, $U(1)_{L_\mu - L_\tau}$, $E_6$ and left-right symmetry. The present analysis is based on detailed Monte Carlo simulations of the expected DUNE-ND signal taking into account detector resolution effects, realistic backgrounds as well as On-Axis and Off-Axis neutrino spectra. We show that the high intensity neutrino beam available at Fermilab can place competitive constraints surpassing those of low-energy neutrino searches and direct detection dark matter experiments.
\end{abstract}

\maketitle

\section{Introduction}

Neutrinos are currently the least understood among all particles in the Standard Model (SM). Contrary to the SM predictions, they are massive and mix with each other as suggested by the discovery of neutrino oscillations in propagation~\cite{deSalas:2020pgw}. Unambiguously, this phenomenon stands out as the clearest evidence of neutrino physics beyond the SM, prompting intense efforts to underpin the ultimate origin of neutrino mass generation~\cite{Schechter:1980gr}. On the theoretical side, massive neutrinos are predicted in a wide class of theoretical models beyond the SM, ranging from simple scalar or fermionic extensions to extended gauge symmetries and Grand Unified Theories (GUT), see e.g. Ref.~\cite{Mohapatra:2005wg}.  Another common feature in such models is the existence of new intermediate  gauge bosons that lead to modifications of the electroweak gauge structure~\cite{Gonzalez-Garcia:1990thl,Gonzalez-Garcia:1990woe}.  Models with extra $U(1)$ symmetry are motivated SM extensions, accommodating dark matter and massive neutrinos~\cite{Das:2017flq,Das:2022oyx}.   Heavy vector bosons~\cite{Langacker:2008yv} have been extensively searched for in the light of LHC data~\cite{ATLAS:2019erb,CMS:2021ctt}. They often lead to charged lepton flavor violating processes~\cite{Deppisch:2013cya,Das:2012ii} as well as they induce novel neutrino interactions~\cite{Miranda:1997vs,Garces:2011aa}.
New physics interactions involving massive neutrinos and new mediators may lead to significant alterations in astrophysical phenomena such as the evolution and cooling mechanism of massive stars~\cite{Raffelt:1996wa} as well as in a modification of the recorded signal at highly sensitive terrestrial experiments~\cite{Giunti:2007ry,Valle:2015pba}. Here we focus on the latter case. In particular we are interested to explore the attainable sensitivities of the Deep Underground Neutrino Detector (DUNE)~\cite{DUNE:2016hlj,DUNE:2022aul} on various new physics scenarios using elastic neutrino electron scattering (E$\nu$ES) events expected to be measured at the Near Detector (ND) facility.

The  Liquid Argon detector i.e. the main component of the 75 ton DUNE-ND will be exposed to the world's most intense beam of high-energy $\nu_e$, 
$\bar{\nu}_e$, $\nu_\mu$, $\bar{\nu}_\mu$ produced by  the Long Baseline Neutrino Facility (LBNF)~\cite{DUNE:2016evb} at Fermilab.  It has been recently pointed out that \eves measurements will be a valuable tool for the determination of the neutrino flux at the GeV scale~\cite{MINERvA:2015nqi}.  For the DUNE setup in particular, through \eves measurements the flux normalization uncertainty can be reduced down to 2\%, while the flux shape uncertainty can be reduced significantly~\cite{Marshall:2019vdy}.  Moreover, the anticipated large exposure combined with the significantly improved background rejection capabilities, makes the ND complex to be a favorable facility for probing several physics opportunities~\cite{DUNE:2020fgq}. This unique experimental setup has motivated a plethora of studies which considered various attractive physics scenarios within and beyond the SM. For instance, Ref.~\cite{deGouvea:2019wav} explored the possibility to determine the weak mixing angle away from the $Z^0$-pole and found that a 2\% precision is possible. Massive neutrinos imply that non-trivial electromagnetic neutrino interactions are possible and various such aspects were examined in Refs.~\cite{Mathur:2021trm,Schwetz:2020xra,Ovchynnikov:2022rqj}. Moreover, the existence of novel vector-type bosons  was analyzed in a series of recent works in the framework of different promising scenarios in which the new particles are contributing to \eves \cite{Altmannshofer:2019zhy,Ballett:2019xoj,Chakraborty:2021apc,Chauhan:2022iuh} or produced via meson decays before in turn they decay to SM particles,  see e.g. Refs.~\cite{Berryman:2019dme,Dev:2021qjj,Capozzi:2021nmp}.  Finally, interesting scenarios leading to sub-GeV dark matter production via dark photon decays were explored in Refs.~\cite{DeRomeri:2019kic,Breitbach:2021gvv}, while a proposal for axion-like particles searches at the DUNE-ND was given in Ref.~\cite{Bhattarai:2022mue}. 

In this work, we first focus on the most general exotic neutrino interactions contributing to E$\nu$ES and  examine their impact in the detectable signal at the DUNE-ND. In particular, we analyze the new physics effects that may occur in the presence of novel mediators predicted in the framework of neutrino generalized interactions (NGI)~\cite{AristizabalSierra:2018eqm,Rodejohann:2017vup}. Thus, all possible Lorentz invariant interactions are taken into account in a model independent way~\cite{Kayser:1979mj}. Contrary to Ref.~\cite{Bischer:2018zcz} which focused on NGIs with heavy mediators, here we consider light mediators which allows us to explore the explicit dependence of the mediator mass for the given NGI in question. Let us also note that while numerous studies analyzed  light mediators of vector-type previously (see e.g. Refs.~\cite{Altmannshofer:2019zhy,Ballett:2019xoj,Chauhan:2022iuh}), in this work the axial-vector, scalar, pseudoscalar and tensor NGI contributions are examined for the first time. We then proceed our analysis by considering motivated scenarios leading to vector-type NGIs such as the $U(1)_{B-L}$ and $U(1)_{L_\mu - L_\tau}$ gauge symmetries, as recently done in Ref.~\cite{Chakraborty:2021apc}. Going one step further, the left-right (LR) symmetric model~\cite{Mohapatra:1980yp,Hati:2017aez} as well as the different realizations of the string-inspired $E_6$ symmetry~\cite{Valle:1990pka} are taken into consideration for the first time in the present work. For all the aforementioned cases we obtain the projected sensitivities of the DUNE-ND by paying special attention in performing realistic simulations of the detected signal taking into account  systematic effects as well as realistic backgrounds. We furthermore compare our extracted sensitivities with existing ones stemming from available experimental data from E$\nu$ES (TEXONO~\cite{TEXONO:2009knm}) and coherent elastic neutrino nucleus scattering--often called CE$\nu$NS--(COHERENT~\cite{COHERENT:2017ipa,COHERENT:2021xmm,COHERENT:2020iec} and Dresden-II~\cite{Colaresi:2022obx}) measurements as well as from solar neutrino experiments (Borexino~\cite{Borexino:2017fbd}),  direct dark matter detection experiments (XENONnT~\cite{XENON:2022ltv} and LZ~\cite{LZ:2022ufs}) and high-energy colliders.

The remainder of the paper has been organized as follows. First, in Sec.~\ref{sec:theory} we describe the new physics scenarios considered in the present work and define the respective \eves cross sections within and beyond the SM. Then, in Sec.~\ref{sec:statistical_analysis} we describe our main procedure for simulating the E$\nu$ES-induced signal at the DUNE-ND as well as our strategy for the extraction of sensitivities. Whenever necessary we also discuss the additional experiments taken into account. Next, in Sec.~\ref{sec:results} we discuss the results obtained in the present study, while in Sec.~\ref{sec:conclusions} we summarize our main conclusions. Additional details are given in the Appendix.

\section{Theoretical framework}
\label{sec:theory}

We now proceed  by introducing the various \eves interactions channels explored in the present work, for which the corresponding cross sections are given and the relevant model parameters are defined. After a brief description of the  well-known SM case our discussion will be mainly focused on the new physics contributions to \eves predicted in the framework of NGIs with light mediators. We then turn our attention to the implications of the heavy vector mediators predicted in the framework of unification models such as the $E_6$ and left-right symmetry.

\subsection{E$\nu$ES through SM interaction Channel}
\label{subsec:EveS_SM}
Within the context of the SM, E$\nu$ES is a well-understood weak interaction process with a tree-level  differential cross section given by~\cite{Kayser:1979mj}
\begin{align}
\label{equn:EveS_SM_xsec_for_nu-alpha}
\begin{split}
\left[\frac{d\sigma_{\nu_\alpha}}{d T_e}\right]_{SM} =&\frac{G_F^2m_e}{2\pi}[(g_V + g_A)^2 +(g_V - g_A)^2\left(1-\frac{T_e}{E_\nu}\right)^2 -(g_V^2-g_A^2)\frac{m_e T_e}{E_\nu^2}] \, ,
\end{split}
\end{align}
where $G_F$ is the Fermi constant, $E_\nu$ is the incoming neutrino energy, while $m_e$ and $T_e$ denote the electron mass and recoil energy, respectively. Here, $g_V$ and $g_A$ denote the flavor-dependent  vector and axial-vector couplings
\begin{equation}
 g_V=-\frac{1}{2}+2\sin^2\theta_W + \delta_{\alpha e}, \qquad  g_A=-\frac{1}{2}+ \delta_{\alpha e} \, ,
 \label{eq:SM_couplings}
\end{equation}
with $\theta_W $ being the weak mixing angle for which we adopt the PDG value $\sin^2 \theta_W =0.23857$ as obtained in the $\overline{\mathrm{MS}}$ renormalization scheme~\cite{ParticleDataGroup:2022pth}. It is important to note that $\nu_e$--$e^{-}$ interactions receive contributions from both neutral-current and charged-current interactions, while $\nu_{\mu, \tau}$--$e^{-}$ interact via the neutral-current only. The Kronecker delta $\delta_{ae}$ in the definition of $g_V$ accounts for this fact.  For the case of antineutrino scattering, the corresponding cross section is given by Eq.(\ref{equn:EveS_SM_xsec_for_nu-alpha}) with the substitution $g_A \to - g_A$. Finally let us stress that for the few GeV neutrino energy accessible  at DUNE, radiative corrections~\cite{Tomalak:2020zfh} amount to only few per mille corrections and hence they can be safely ignored, thus leaving the weak mixing angle as the only source of theoretical uncertainty.

\subsection{E$\nu$ES through light novel mediators}
\label{subsec:EveS_BSM}
 Neutrino nonstandard interactions (NSI) has become the subject of extensive research using both low-~\cite{Miranda:2015dra} and high-energy~\cite{Babu:2020nna} neutrino scattering  as well as  direct dark matter detection~\cite{Amaral:2023tbs} and oscillation~\cite{Farzan:2017xzy} data (for motivations, UV complete models and several applications of NSI see e.g.~\cite{Proceedings:2019qno}). While NSIs usually assume the existence of a new vector (or axial-vector) type mediator, the NGI framework~\footnote{The complete operator basis that includes also couplings to gluons and photons with all possible operators up to dimension 7 has been presented in the context of Effective Field Theory in Ref.~\cite{Altmannshofer:2018xyo}.} considered in the present work accommodates a wider class of interactions below the electroweak symmetry breaking scale that arise from the Lagrangian~\cite{Rosen:1982pj,Rodejohann:2017vup}
\begin{equation}
\mathscr{L}_\text{NGI}=\frac{G_F}{\sqrt{2}}\sum_{\substack{X=S, P, V, A, T \\ \alpha=e,\mu,\tau}}C_{\alpha,\alpha}^{f,P}\left[\bar{\nu}_\alpha\Gamma^X L \nu_\alpha\right]\left[\bar{f}\Gamma_X P f\right] \, .
 \label{equn:CEvNS_BSM_Effective_Lagrangian}
 \end{equation}
In the above Lagrangian, $\Gamma_X= \{\mathbb{1}, i\gamma_5, \gamma_\mu, \gamma_\mu\gamma_5, \sigma_{\mu\nu} \}$ (with $\sigma_{\mu\nu}=\frac{i}{2}[\gamma_\mu,\gamma_\nu])$ which enables a phenomenological study of all possible Lorentz-invariant structures corresponding to  $ X= \{S, P, V, A, T\}$ interactions in a model-independent way. NGIs lead to corrections in both \cevns (for $f= \{u,d\}$)~\cite{AristizabalSierra:2018eqm,Lindner:2016wff,DeRomeri:2022twg,Majumdar:2022nby} and \eves (for $f=e$)~\cite{Khan:2019jvr,Majumdar:2021vdw,A:2022acy} cross sections. Here we focus on the latter. The dimensionless coefficients quantify the strength of the interaction $X$ with respect to the Fermi constant through the relation $C_{\alpha,\alpha}^{f,P}=(\sqrt{2}/G_F) (g_X^2/(q^2+m_X^2))$,  while $m_X$ and $g_X$ denote the mass and coupling of the respective mediator. Since through \eves or \cevns  measurements only products of couplings  $g_{\nu X} g_{f X}$ can be probed,  for later convenience in our DUNE-based analysis we take the   coupling to be $g_X = \sqrt{g_{\nu X} g_{e X}}$. 

For sufficiently light vector and axial-vector  mediators, the corresponding differential cross sections are taken from the SM one given in Eq.(\ref{equn:EveS_SM_xsec_for_nu-alpha}) and the replacement~\cite{Lindner:2018kjo,Ballett:2019xoj} 
\begin{equation}
\label{equn:EveS_BSM_xsec_V,A}
g'_{V/A}=g_{V/A}+\frac{g_{\nu V/A}\cdot g_{e V/A}}{\sqrt{2}G_F(2m_e T_e + m_{V/A}^2)} \, .
\end{equation}
At this point, it should be stressed that the aforementioned general vector NGIs are not particularly interesting since they do not follow from anomaly-cancellation. We are thus motivated to emphasize their connection to well-known anomaly-free models such as those arising in the framework of extra $U(1)_{B-L}$ or $U(1)_{L_\mu - L_\tau}$ gauge symmetries. Interestingly in the former case,  the differential cross section is identical to the general vector NGI one described above, see e.g. the discussion of Refs.~\cite{AristizabalSierra:2020edu, Majumdar:2021vdw}. In the latter case instead, \eves contributions arise from the kinetic mixing induced at the one loop-level as described in Ref.~\cite{Amaral:2021rzw}. Then, the corresponding vector coupling can be expressed in the form~\cite{Altmannshofer:2019zhy}
\begin{equation}
g'_V=  g_V \pm \frac{\alpha_{em}}{3 \sqrt{2} \pi G_F} \log{\left(\frac{m^2_\tau}{m_\mu^2} \right)} \frac{g_{\nu V}\cdot g_{e V}}{(2m_e T_e + m_{V}^2)} \, ,
\label{equn:EveS_BSM_xsec_LVM_LmLt}
\end{equation}
where the plus (minus) sign accounts for $\nu_\tau$ $(\nu_\mu)$ scattering off electrons, while for $\nu_e$ only SM interactions are allowed. Finally, the scalar, pseudoscalar and tensor cross sections can be cast in the form ~\cite{Khan:2020vaf, Coloma:2022umy, Link:2019pbm}
\begin{equation}
\left[\frac{d\sigma_{\nu_\alpha}}{dT_e}\right]_{S} =\left[\frac{g^2_{\nu S}\cdot g^2_{eS}}{4\pi(2m_e T_e + m_{S}^2)^2}\right]\frac{m_e^2  T_e}{E_\nu^2}   \left(1 + \frac{T_e}{2 m_e} \right) \, ,
\end{equation}
\begin{equation}
\left[\frac{d\sigma_{\nu_\alpha}}{d T_e}\right]_{P} = \left[  \frac{g^2_{\nu P}\cdot g^2_{eP}}{8\pi(2m_e T_e + m_{P}^2)^2} \right]\frac{m_e T_e^2}{E_\nu^2}\, ,
\end{equation}
\begin{equation}
\label{equn:EvES_Tensor}
\left[\frac{d\sigma_{\nu_\alpha}}{d T_e}\right]_{T} =\frac{m_e\cdot g^2_{\nu T}\cdot g^2_{eT}}{\pi(2m_e T_e + m_{T}^2)^2}\cdot\left[1+2\left(1-\frac{T_e}{E_\nu}\right)+\left(1-\frac{T_e}{E_\nu}\right)^2-\frac{m_e T_e}{E_\nu^2}\right]\, .
\end{equation}
We should finally stress that, unlike the $X=V,A$ cases, for $X=S,P,T$  there is absence of interference with the SM cross section.  Note also that for the $\sim$GeV recoil energies involved at DUNE-ND, it holds that $1 \ll \frac{T_e}{2 m_e}$ and therefore the scalar and pseudoscalar cross sections become identical. This will become evident in our main results below~\footnote{For the case of low-energy scattering experiments such as XENONnT, LZ, TEXONO and COHERENT where the recoil energies involved are of the few keV order, it holds that $1 \gg \frac{T_e}{2 m_e}$ and the second term is usually dropped in the literature~\cite{Majumdar:2022nby, A:2022acy,Cerdeno:2016sfi}.}.

\subsection{Left-Right Symmetry}

A  heavy neutral vector boson, $Z^\prime$, occurs in models based in the $SU(2))_L\otimes SU(2)_R\otimes U(1)_{B-L}$ gauge group~\cite{Erler:1999ub}\footnote{ A charged gauge boson arises as well, however it is not relevant for the neutral-current interactions we are interested in the present work.}. As explained in Ref.~\cite{Miranda:1997vs} LR symmetric models are particularly appealing because of their  ability to incorporate parity violation alongside gauge symmetry breaking, rather than requiring manual intervention as in the SM.
Within the context of LR symmetry the  corresponding \eves differential cross sections are obtained from Eq.(\ref{equn:EveS_SM_xsec_for_nu-alpha}) through the 
substitutions $g_L \to f_L^{LR}$ and $g_R \to f_R^{LR}$~\cite{Barranco:2007tz}
\begin{equation}
\begin{aligned}
f_{L}^{L R}= & \mathcal{A} g_{L} + \mathcal{B} g_{R} \, , \\
f_{R}^{L R}=& \mathcal{A} g_{R} + \mathcal{B} g_{L} \, ,
\end{aligned}
\end{equation}
and the definitions
\begin{equation}
\mathcal{A} = 1+\frac{\sin^4 \theta_W}{1-2 \sin^2 \theta_W} \gamma \, , \qquad \mathcal{B} =\frac{\sin^2 \theta_W \left(1-\sin^2 \theta_W\right)}{1-2 \sin^2 \theta_W}\gamma \, ,
\label{eq:LR_params}
\end{equation}
where $\gamma=\left(M_{Z^0} / M_{Z^{\prime}}\right)^{2}$ ($Z^0$ is the SM vector boson).
Note, that the left- and right-handed SM couplings $g_L$ and $g_R$ are related to the vector and axial-vector couplings defined in Eq.(\ref{eq:SM_couplings}) according to
\begin{equation}
g_L = \frac{g_V+g_A}{2} \, , \qquad g_R = \frac{g_V-g_A}{2}.
\end{equation}

\subsection{$E_6$ models}

Turning our attention to GUTs, we are intended to explore potential signatures--detectable at the DUNE-ND via \eves measurements--that could arise from the new interactions predicted in the presence of the primordial $E_6$ gauge symmetry. Being a rank-6 group, $E_6$ yields two novel neutral gauge bosons associated to the two new hypercharges, namely $\chi$ and $\psi$, which follow from the respective extra $U(1)$ symmetries present in $E_6/SO(10)$ and in $SO(10)/SU(5)$, see e.g. Ref.~\cite{Erler:1999ub}. The corresponding  quantum numbers for $Y_\chi$ and $Y_\psi$ are listed in Table~\ref{tab:E6}. As explained in Ref.~\cite{Gonzalez-Garcia:1990woe}, in the low-energy
  regime $E_6$ yields a single $U(1)$ symmetry that is written as a combination of 
  $U(1)_\chi$ and $U(1)_\psi$ symmetries.  Then, a one-parameter family of models is defined with hypercharge 
\begin{equation}
Y_\beta =Y_{\chi}\cos \beta +Y_{\psi}\sin \beta \, , 
\end{equation}
where $ Q=T^3+ Y $ represents the charge operator. 
\begin{table}[t]
   
\begin{tabular}{cccc}
\hline {\rule[-3mm]{0mm}{8mm}} & $T_3$ & $\sqrt{40}Y_\chi$ &
$\sqrt{24}Y_\psi$   \\
\hline
$Q$   & $\begin{pmatrix} 1/2 \\ -1/2 \end{pmatrix}$ & $-1$ & $1$ \\
$u^c$ &       $0$               & $-1$ & $1$ \\
$e^c$ &       $0$               & $-1$ & $1$ \\
$d^c$ &       $0$               & $ 3$ & $1$ \\
$l$   & $ \begin{pmatrix} 1/2 \\ -1/2 \end{pmatrix}$ & $ 3$ & $1$ \\
\hline
\end{tabular}  
\caption{Quantum numbers for the light particles in the {\bf 27} of $E_6$~\cite{Gonzalez-Garcia:1990thl}.}
\label{tab:E6}
\end{table}
The modifications to the SM \eves couplings that follow from the low-energy effective Lagrangian have been previously written as~\cite{Barranco:2007tz}
\begin{equation}
\begin{aligned} 
f_{L} & = g_{L}+ \varepsilon_{L} \, ,\\
f_{R} & = g_{R}+ \varepsilon_{R} \, ,
\end{aligned}
\end{equation}
where the new $E_6$-induced contributions take the form
\begin{equation}
\begin{aligned}
\varepsilon_{L}=& 2 \gamma \sin ^{2} \theta_{W} \left(\frac{3 c_{\beta}}{2 \sqrt{6}}+\frac{s_{\beta}}{3} \sqrt{\frac{5}{8}}\right)^{2} \, , \\ 
\varepsilon_{R}=& 2 \gamma \sin ^{2} \theta_{W} \left(\frac{c_{\beta}}{2 \sqrt{6}}-\frac{s_{\beta}}{3} \sqrt{\frac{5}{8}}\right)\left(\frac{3 c_{\beta}}{2\sqrt{6}}+\frac{s_{\beta}}{3} \sqrt{\frac{5}{8}}\right) \, ,
\end{aligned}
\end{equation} 
while $\gamma$ is defined as previously and the abbreviations $c_{\beta}=\cos \beta$, $s_{\beta}=\sin \beta$ have been used. While any value of $\cos \beta$ is allowed, here we will focus on the three most notable $E_6$ models i.e., the $(\chi, \psi, \eta )$ model with    $\cos \beta=(1, 0, \sqrt{3/8})$~\cite{Gonzalez-Garcia:1990thl}. Notice that for $\cos \beta = - \sqrt{5/32}$ a full cancellation occurs and \eves is not sensitive to $E_6$ models for this particular case.

\section{Statistical analysis}
\label{sec:statistical_analysis}

\subsection{Simulation of \eves signal at the DUNE-ND}

\begin{figure}[ht]
\captionsetup{justification=centering}
\subcaptionbox{Small mediator mass}
{\includegraphics[width = \textwidth]{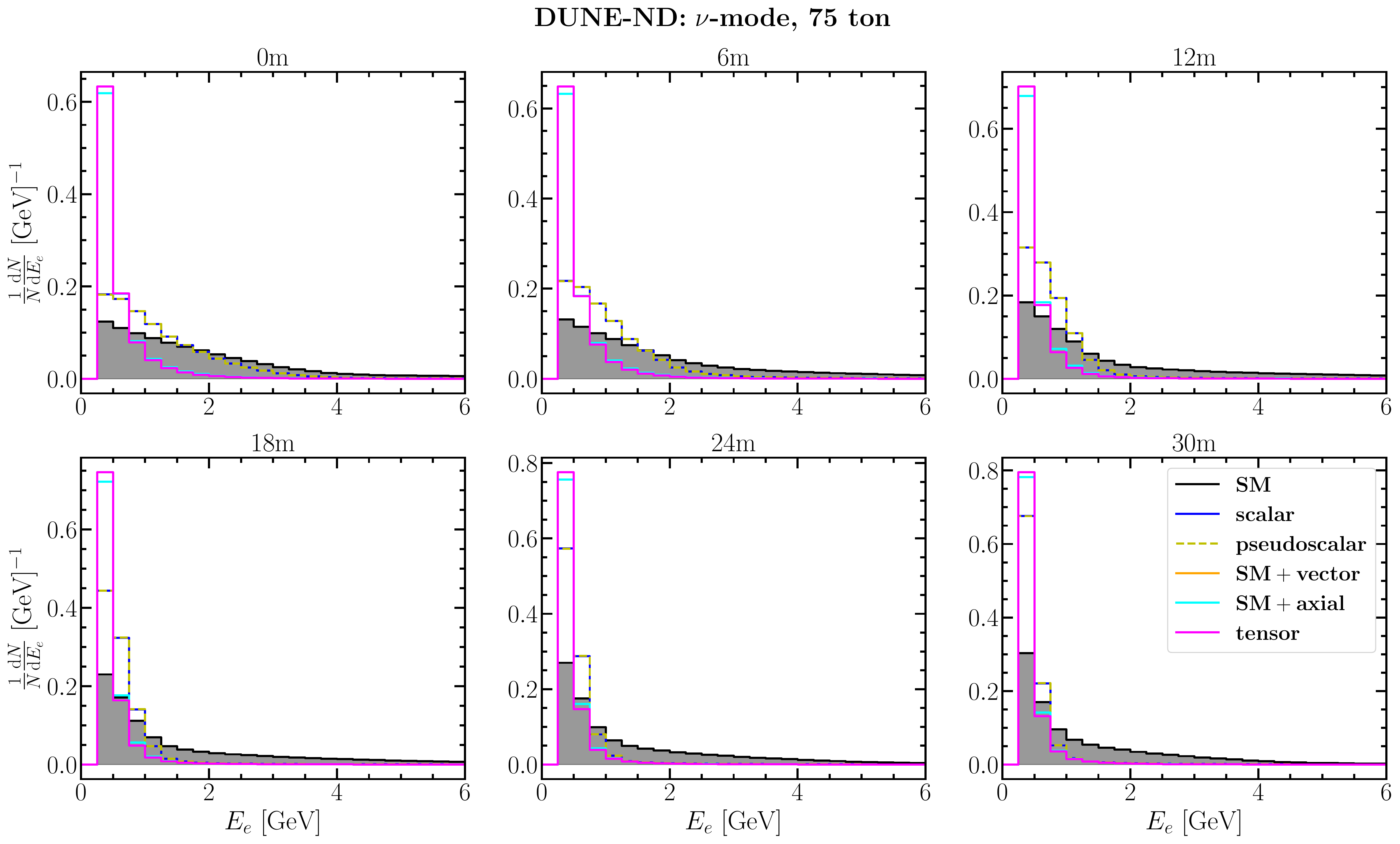}}
\captionsetup{justification=centering}
\subcaptionbox{Large Mediator mass}
{\includegraphics[width = \textwidth]{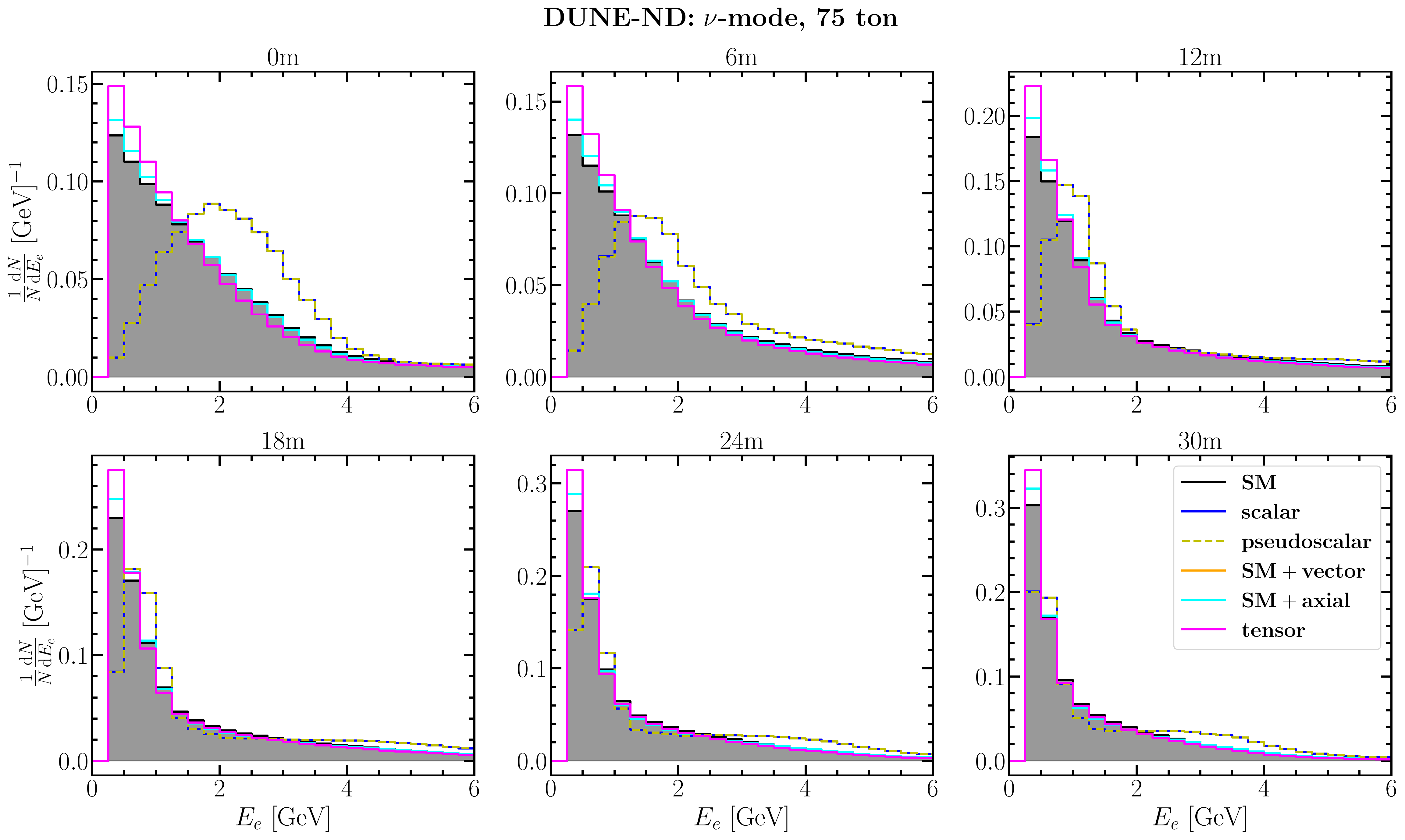}}
\caption{Normalized differential event rates of the different interactions $X=S,P,V,A,T$ for the On-Axis and the various Off-Axis locations, assuming the neutrino-mode operation of DUNE. The upper panel (a) illustrates the case of small mediator mass $m_X=10$~MeV while the lower panel the case of large mediator mass $m_X=1$~GeV.}
\label{fig:diff_rates_neutrino}
\end{figure}

\begin{figure}[t]
\captionsetup{justification=centering}
\subcaptionbox{Small mediator mass}
{\includegraphics[width = \textwidth]{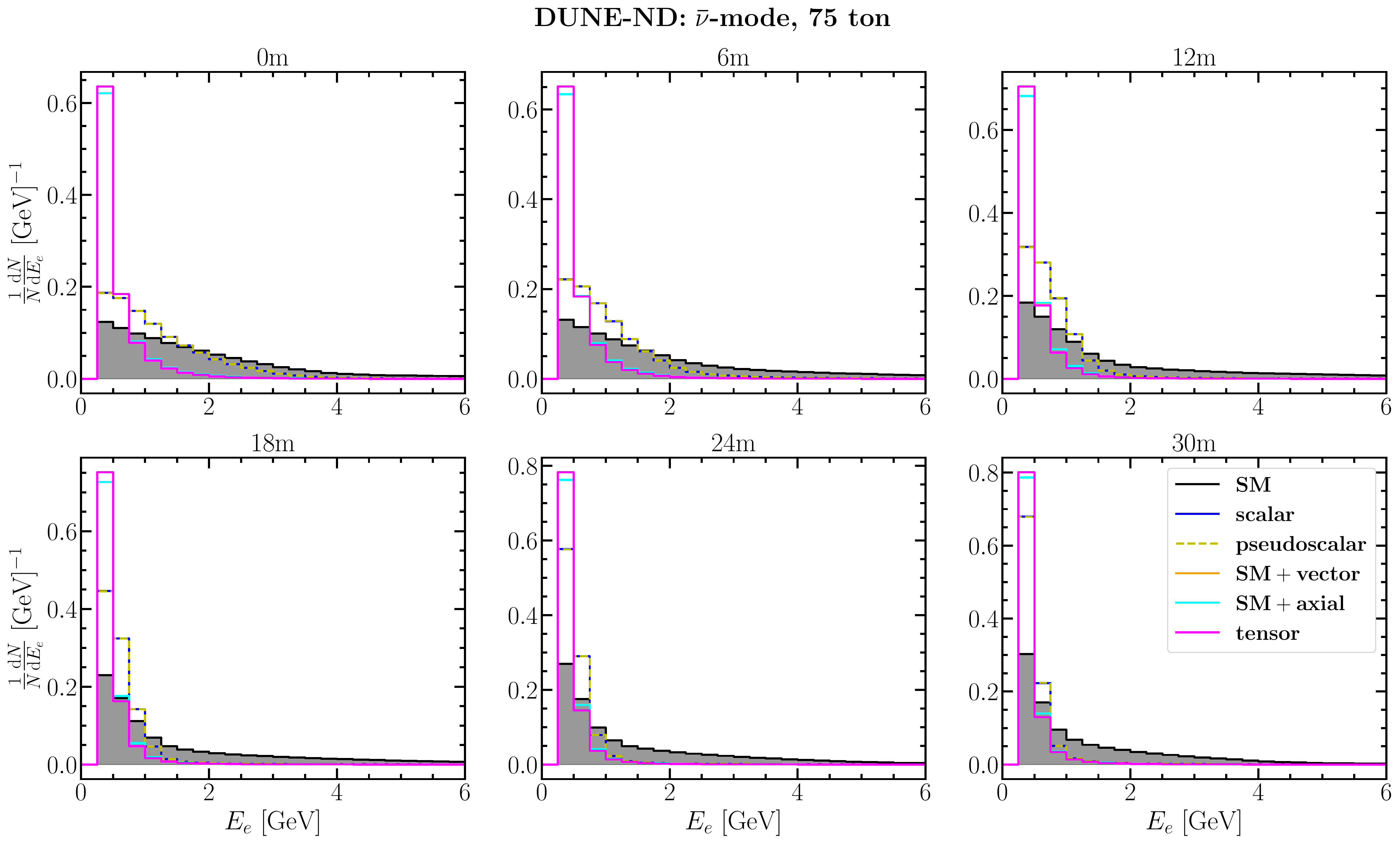}}
\captionsetup{justification=centering}
\subcaptionbox{Large Mediator mass}
{\includegraphics[width = \textwidth]{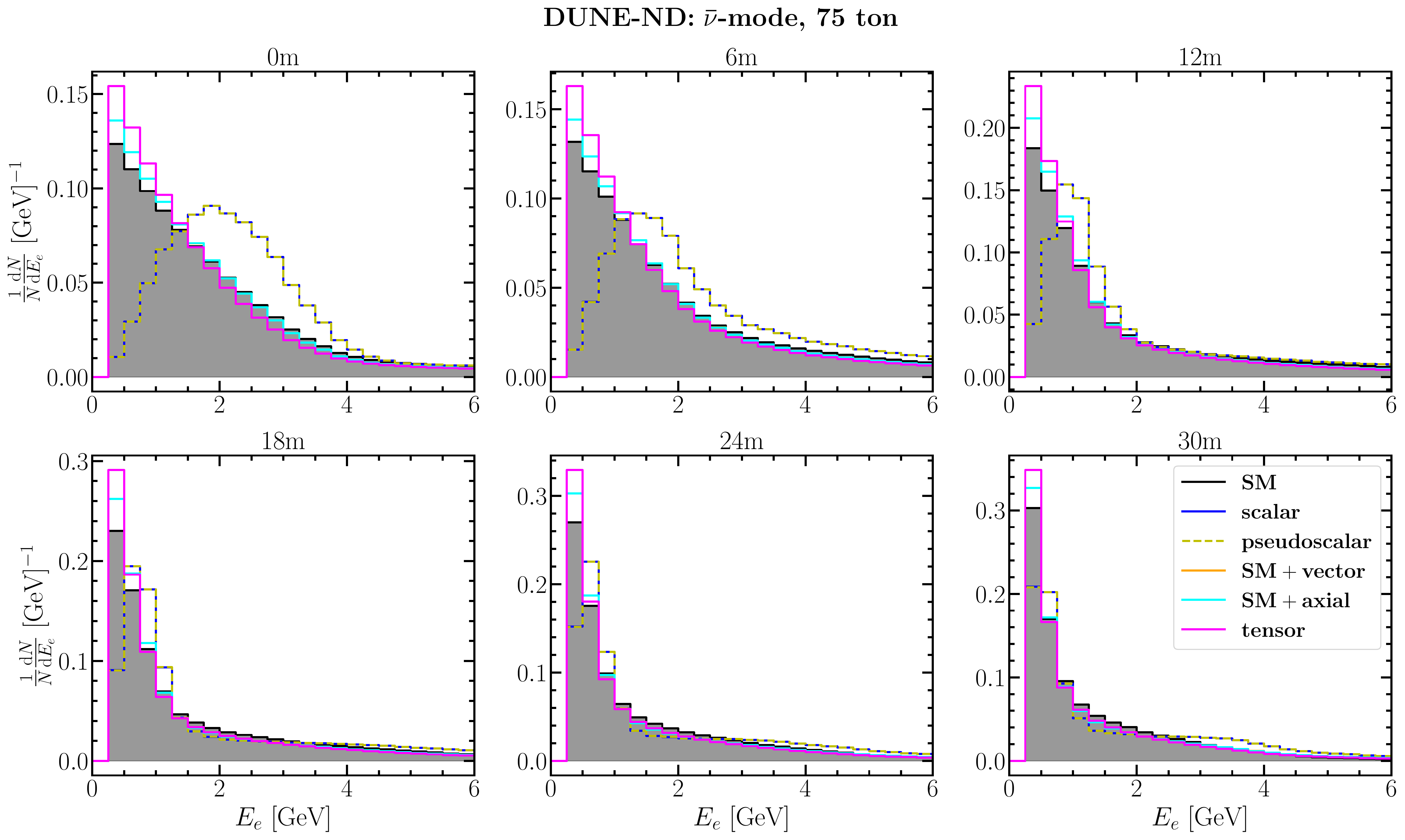}}
\caption{Same as Fig.~\ref{fig:diff_rates_neutrino} but for the antineutrino mode.}
\label{fig:diff_rates_antineutrino}
\end{figure}

Neglecting acceptance and resolution effects (see the discussion below) the expected event rates with respect to the electron recoil energy at DUNE-ND can be evaluated from the expression
\begin{equation}
\left[\frac{dN}{d T_e }\right]_\lambda =t_\mathrm{run}  N_e N_\mathrm{POT}\sum_{\alpha}\int_{E_\nu^\mathrm{min}}^{E_\nu^\mathrm{max}}dE_{\nu}\frac{d\Phi_{\nu_\alpha}(E_\nu)}{dE_\nu}  \left[\frac{d\sigma_{\nu_{\alpha}}}{d T_e }\right]_\lambda  \,,
\label{equn:EvES_Differential_Event_Rate}
\end{equation}
where $t_\mathrm{run}$ and $N_e$ stand for the total running period of the experiment and the number of electron targets at the 75 ton $^{40}\text{Ar}$ DUNE-ND, while $\frac{d\Phi_{\nu_\alpha}(E_\nu)}{dE_\nu}$ is the corresponding incoming neutrino flux for the different flavors $\nu_\alpha$ taken from~\cite{Fields:2017}. Here $N_\mathrm{POT}=1.1 \times 20^{21}$ denotes  the number of protons on target (POT) per year assuming a 120 GeV proton beam~\cite{DUNE:2016hlj}.  From the plot it becomes evident that the expected scalar and pseudoscalar signals are indistinguishable (see the discussion in Subsec.~\ref{subsec:EveS_BSM}). The lower integration limit is trivially obtained from the kinematics of the process and reads $E_\nu^\mathrm{min} = \left( T_e + \sqrt{T_e^2 + 2 m_e T_e} \right)/2$, while the upper limit corresponds to the endpoint of the incoming neutrino energy distribution. Finally, $\lambda$ accounts for SM interactions as well as the various new physics interactions predicted within the context of NGI, or $E_6$ and LR symmetries. Following Ref.~\cite{deGouvea:2019wav}, we consider a recoil energy threshold $T_e^\mathrm{th}=50$~MeV and restrict our analysis in the range $T_e^\mathrm{th} \leq T_e \leq T_e^\mathrm{max}$ with $T_e^\mathrm{max}=20$~GeV. For the various On-Axis and Off-Axis locations, example spectra as a function of the total electron energy $E_e = T_e + m_e$ are shown in Figs.~\ref{fig:diff_rates_neutrino} and \ref{fig:diff_rates_antineutrino} which correspond to the neutrino and antineutrino modes, respectively. For the sake of comparison between the different NGIs $X=\{S,P,V,A,T\}$, the differential event rates are presented as normalized distributions in order to make evident the various spectral shapes. Moreover, the upper (lower) panels are plotted assuming a small (large) mediator mass of $m_X=10~\mathrm{MeV}$ ($m_X=1~\mathrm{GeV}$) from where a significant dependence on the mediator mass is found.

As recently pointed out in Ref.~\cite{DeRomeri:2019kic}, the main backgrounds coming from charged-current quasielastic (CCQE) neutrino scattering on LArTPC and $\pi^0$ missidentification can be vetoed by a cut on $E_e \theta_e^2$~\cite{MINERvA:2015nqi, Marshall:2019vdy}, where $\theta_e$ denotes the scattering angle.  Therefore, in what follows we choose to express the \eves signal in terms of the quantity $E_e \theta_e^2$.  From the kinematics of the process one has $1 - \cos \theta_e = m_e \frac{1-y}{E_e}$, where $y= T_e/E_\nu$ denotes the inelasticity which takes values in the range $T_e^\mathrm{th}/E_\nu \leq y \leq 1$. Since \eves is forward peaked, the electron recoil energy can be expressed in terms of $E_e \theta_e^2$ as $T_e = E_\nu \left(1 - \frac{E_e \theta_e^2}{2 m_e} \right)$ with  $E_e \theta_e^2 < 2 m_e$ being an upper limit imposed by the kinematics of the process. Therefore, for the DUNE-ND it is more convenient to evaluate the expected neutrino signal in $E_e \theta_e^2$ space  through the expression
\begin{equation}
\left[\frac{dN}{d E_e \theta_e^2}\right]_\lambda =t_{run}  N_e N_\mathrm{POT} \sum_{\alpha}\int_{E_\nu^\mathrm{min}}^{E_\nu^\mathrm{max}}dE_{\nu}\frac{d\Phi_{\nu_\alpha}(E_\nu)}{dE_\nu}  \left[\frac{d\sigma_{\nu_{\alpha}}}{d E_e \theta_e^2 }\right]_\lambda  \,,
\label{equn:EvES_Differential_Event_Rate_EeThetaw}
\end{equation}
where 
\begin{equation}
\frac{d\sigma_{\nu_{\alpha}}}{d E_e \theta_e^2 } = \frac{E_\nu}{2 m_e}\frac{d\sigma_{\nu_{\alpha}}}{d T_e } \Big\vert_{T_e = E_\nu \left(1 - \frac{E_e \theta_e^2}{2 m_e} \right)}\, .
\end{equation}
We have verified that the effect of energy resolution $\sigma_{E_e}/E_e = 10\% / \sqrt{E_e/\mathrm{GeV}}$ considered in Refs.~\cite{deGouvea:2019wav,Mathur:2021trm} is negligible, in agreement with Ref.~\cite{Chakraborty:2021apc}, thus in our analysis only the effect of angular resolution will be considered.  On the other hand, regarding the scattering angle taking a usual Gaussian function can overestimate the smearing of the angular resolution, since the azimuthal angle is important when applying the smearing.  For this reason, the finite angular resolution of DUNE-ND is taken into account by following closely the procedure of Ref.~\cite{Mathur:2021trm}. Taking the neutrino beam to be in the $\hat{z}$ direction and letting the true scattering angle to be $\theta_e^t$, the true electron momentum vector is simply written as ${\hat{p}}^t_e=(\sin\theta_e^t ~\hat{x},0,\cos\theta_e^t~\hat{z})$. The latter is related to the reconstructed momentum vector through the relation~\cite{deGouvea:2019wav}
\begin{equation}
    \hat{p}_e^\text{reco}=R_{\hat{y}}(\theta_t)R_{\hat{z}}(\phi_2)R_{\hat{y}}(\theta_1)R_{\hat{y}}(-\theta_e^t){\hat{p}}^t_e\,,
    \label{eq:reco_momentum}
\end{equation}
where $R_{\hat{i}}(\alpha)$ denotes the rotation matrix about the axis $\hat{i}$ through the angle $\alpha$, while the reconstructed angle $\theta_e^\text{reco}$ is given by $\theta_e^\text{reco}=\cos^{-1}(\hat{z} \cdot \hat{p}_e^\text{reco})$. The rotations are performed in order to have control on the smearing. In particular, $R_{\hat{y}}(-\theta_e^t)$ rotates the electron momentum in the $z$-axis. The next two rotations are performed to apply the smearing, first around the $y$-axis to modify the polar angle, and the second around the $z$-axis to modify the azimuth. The final rotation is performed to undo the first rotation. 

We calculate the reconstructed event spectrum based on the following procedure. First, we choose randomly the electron recoil $T_e$ and initial neutrino $E_\nu$ energies in a given bin of $(E_\nu, T_e)$ from which we determine the values of the electron energy $E_e$ and the true polar angle theta $\theta_e^t$. Then, we calculate the true $(E_e \theta_e^2)^t$ \eves event spectrum that corresponds to the given $(E_\nu, T_e)$ bin using Monte Carlo integration. For each bin, we furthermore evaluate the reconstructed electron momentum by performing the rotations given in Eq.(\ref{eq:reco_momentum}) as follows: for each $\theta_e^t$, we choose a new angle $\theta_1$ from a Gaussian distribution with $\sigma_\theta$ being the angular resolution, while a new angle $\phi_2$  is randomly chosen from a uniform distribution within $[0, 2\pi]$. Then, we obtain the reconstructed angle $\theta_e^\text{reco}$ from which we construct the $(E_e \theta_e^2)^\text{reco}$ event spectrum in the given $(E_\nu, T_e)$ bin. We finally repeat the procedure for the other bins and sum up the individual event spectra.

\begin{figure}[t]
\includegraphics[width = 0.32 \textwidth]{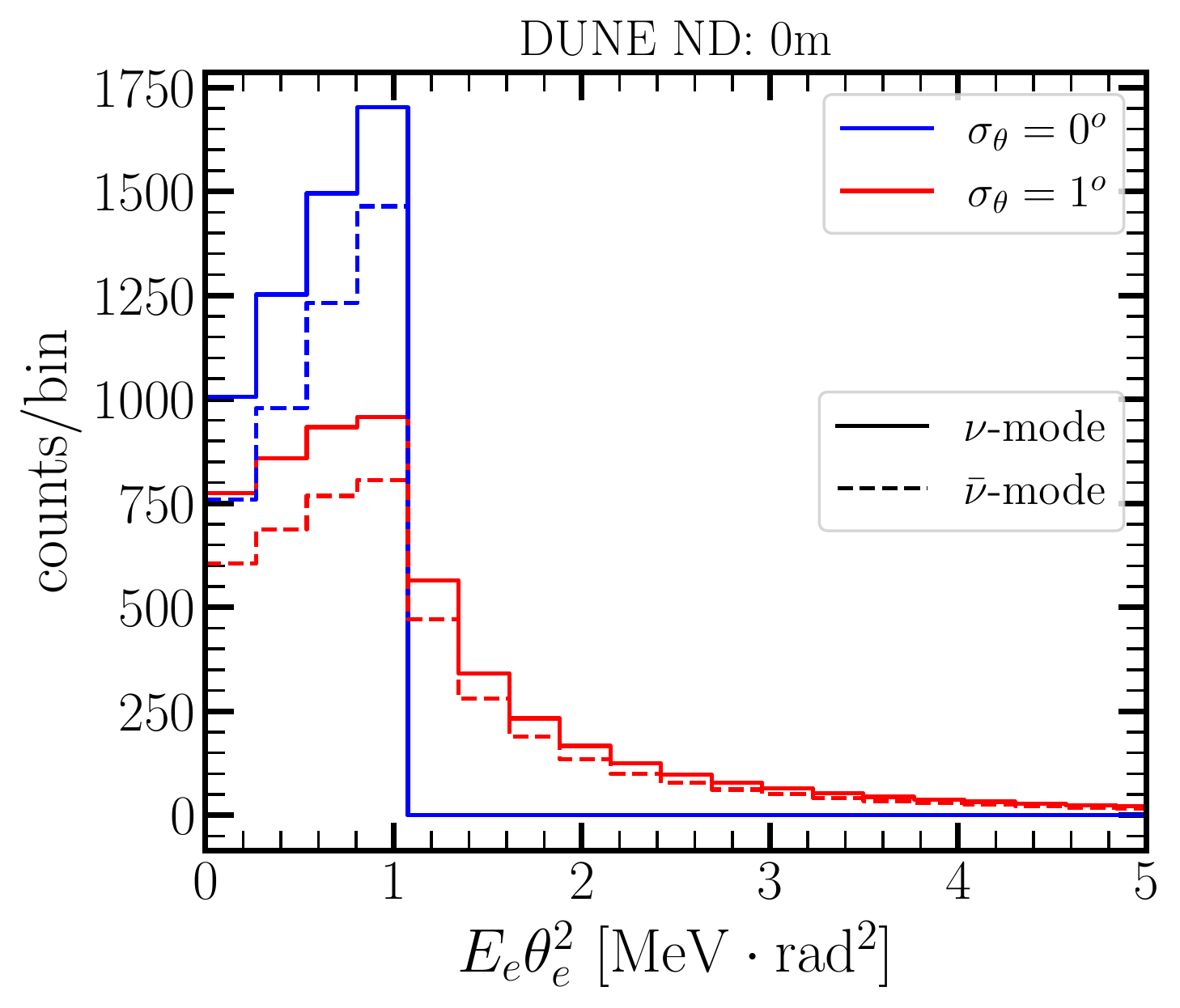}
\includegraphics[width = 0.32 \textwidth]{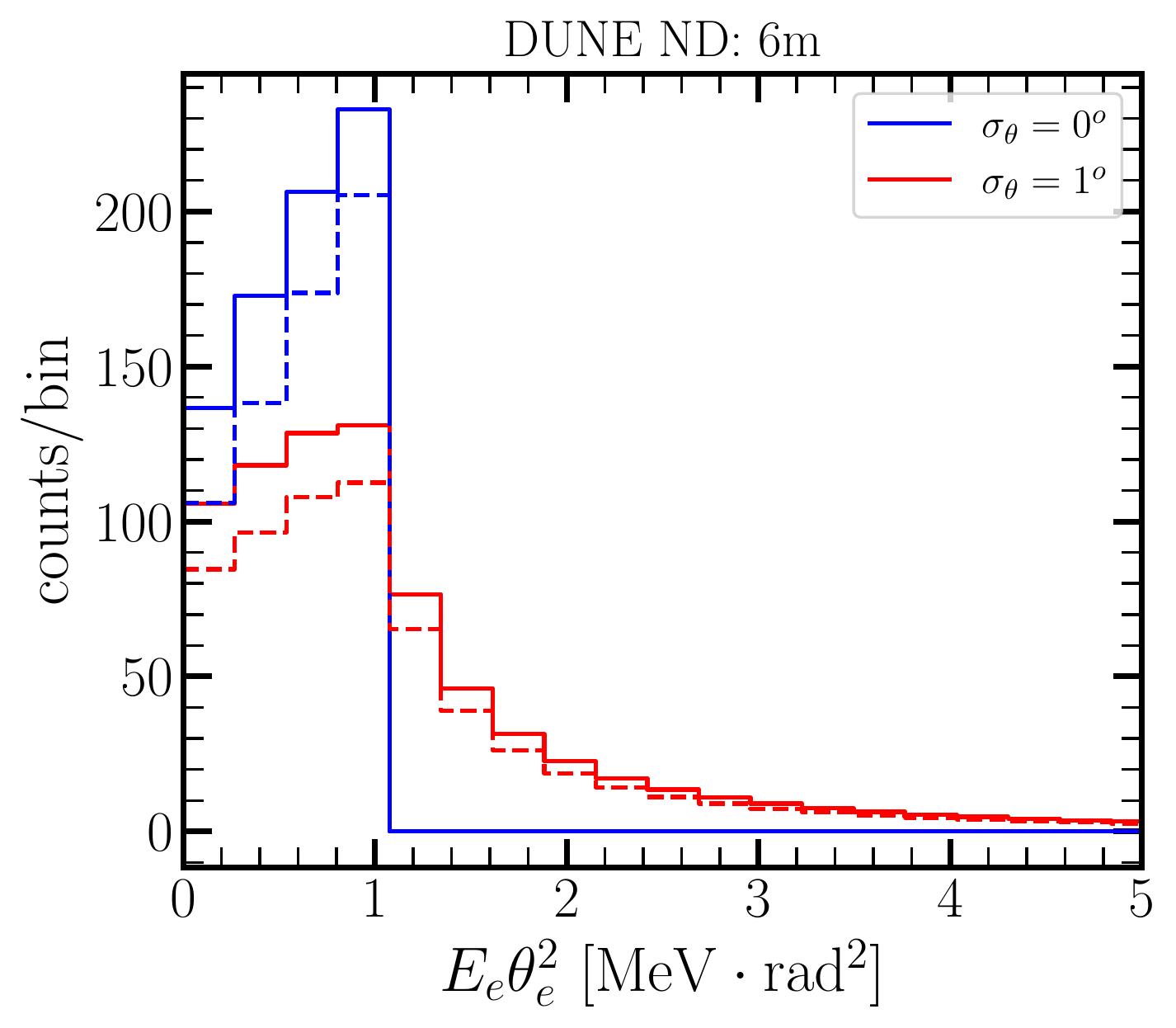}
\includegraphics[width = 0.32 \textwidth]{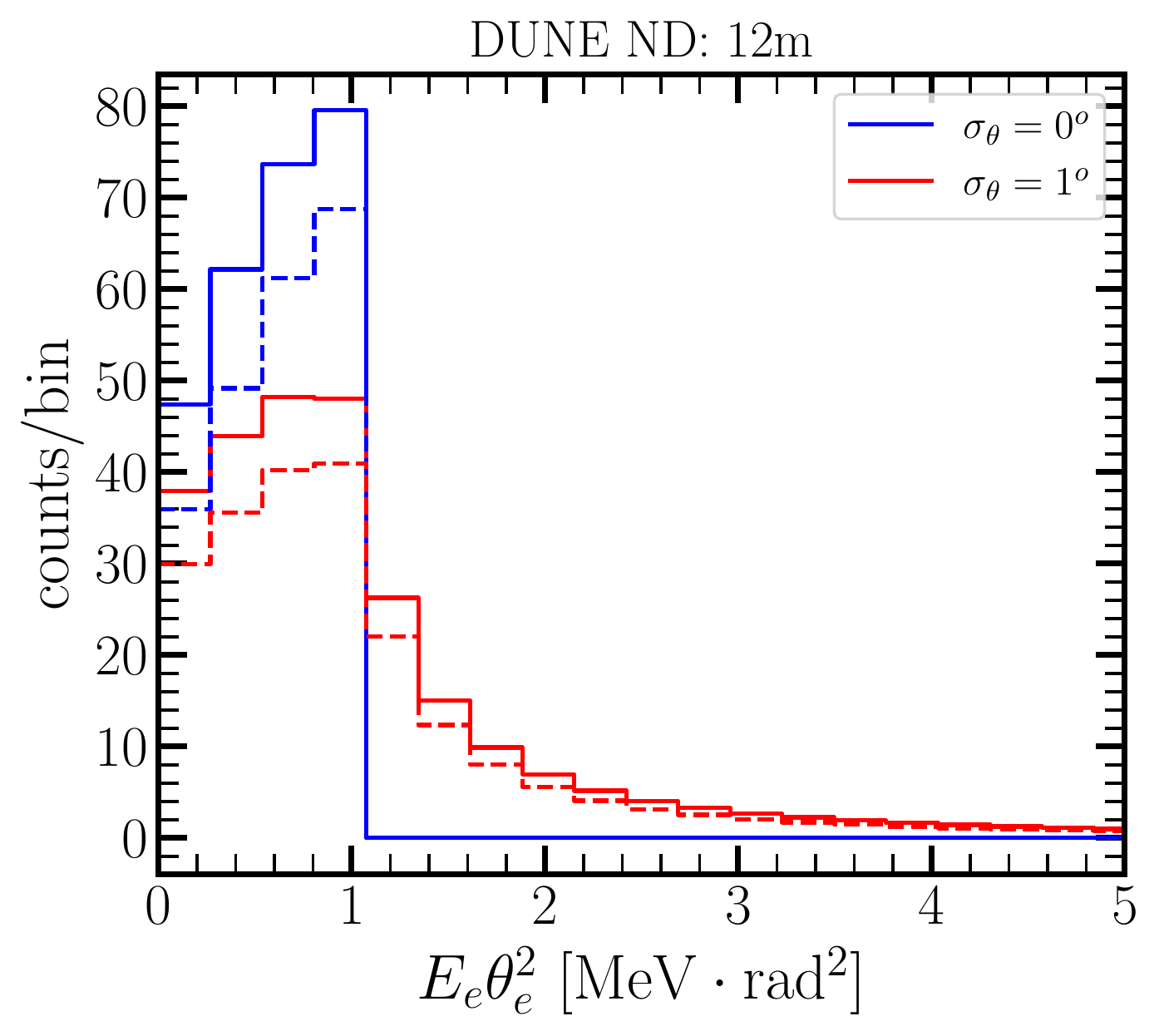}
\includegraphics[width = 0.32 \textwidth]{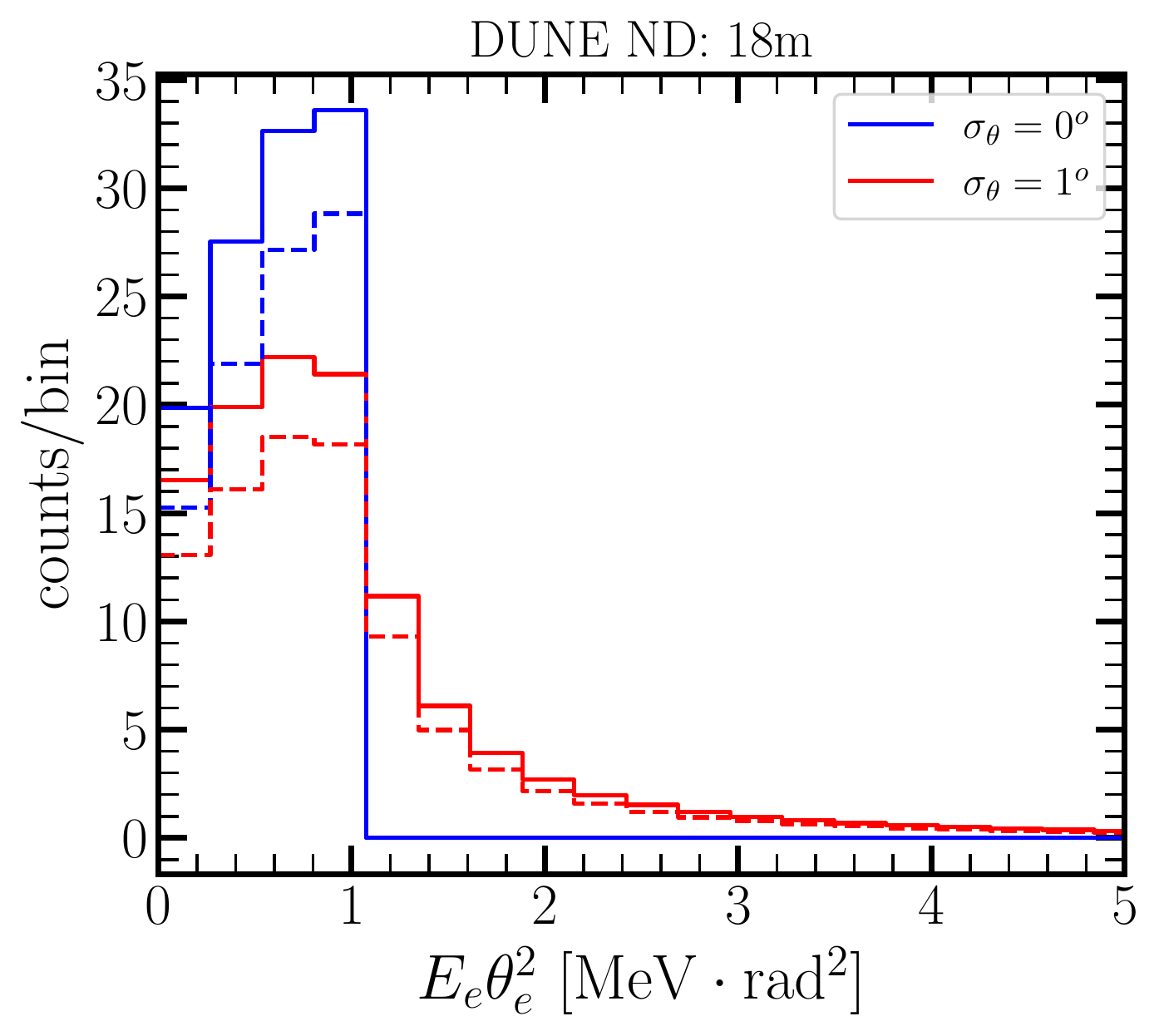}
\includegraphics[width = 0.32 \textwidth]{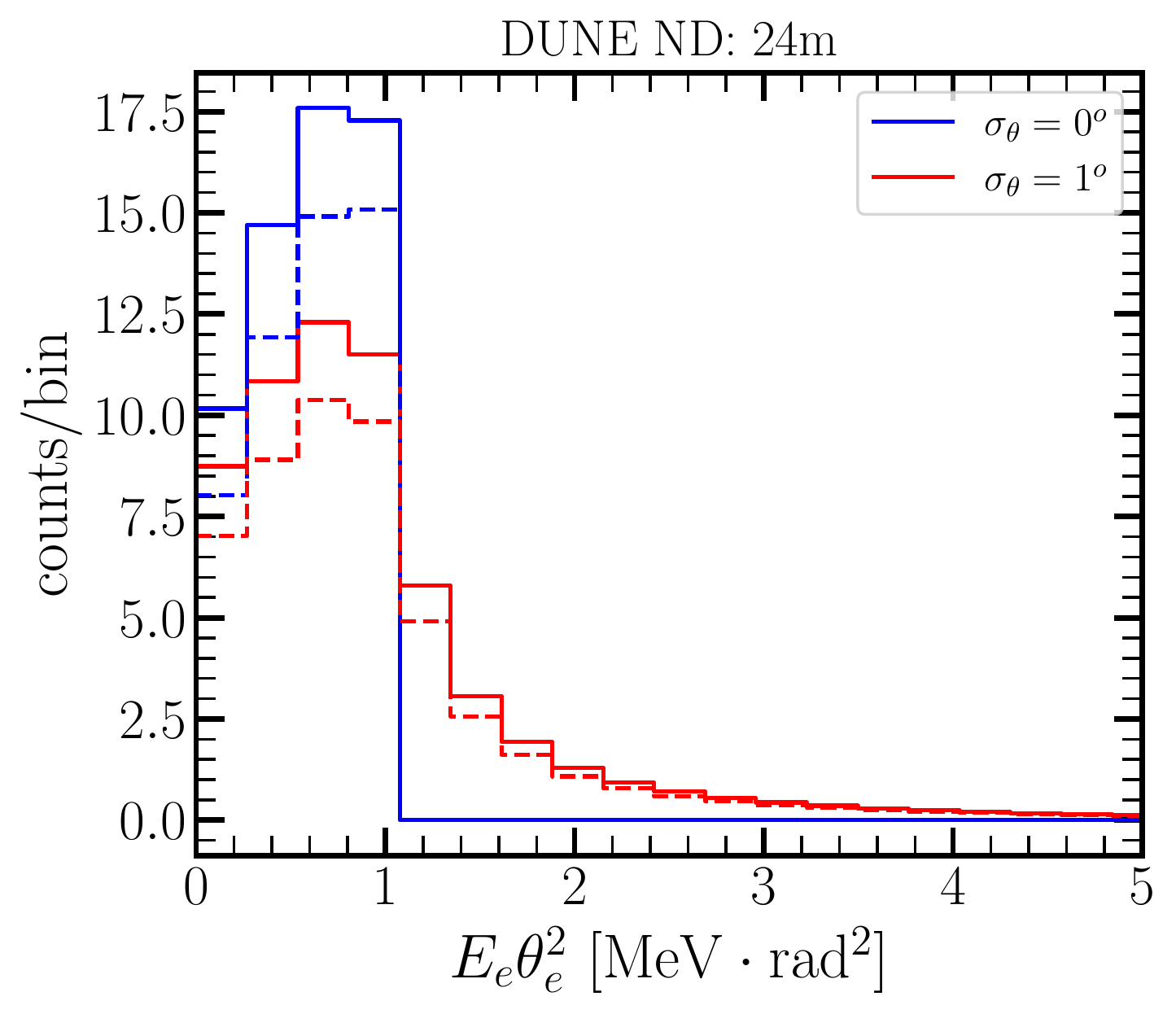}
\includegraphics[width = 0.32 \textwidth]{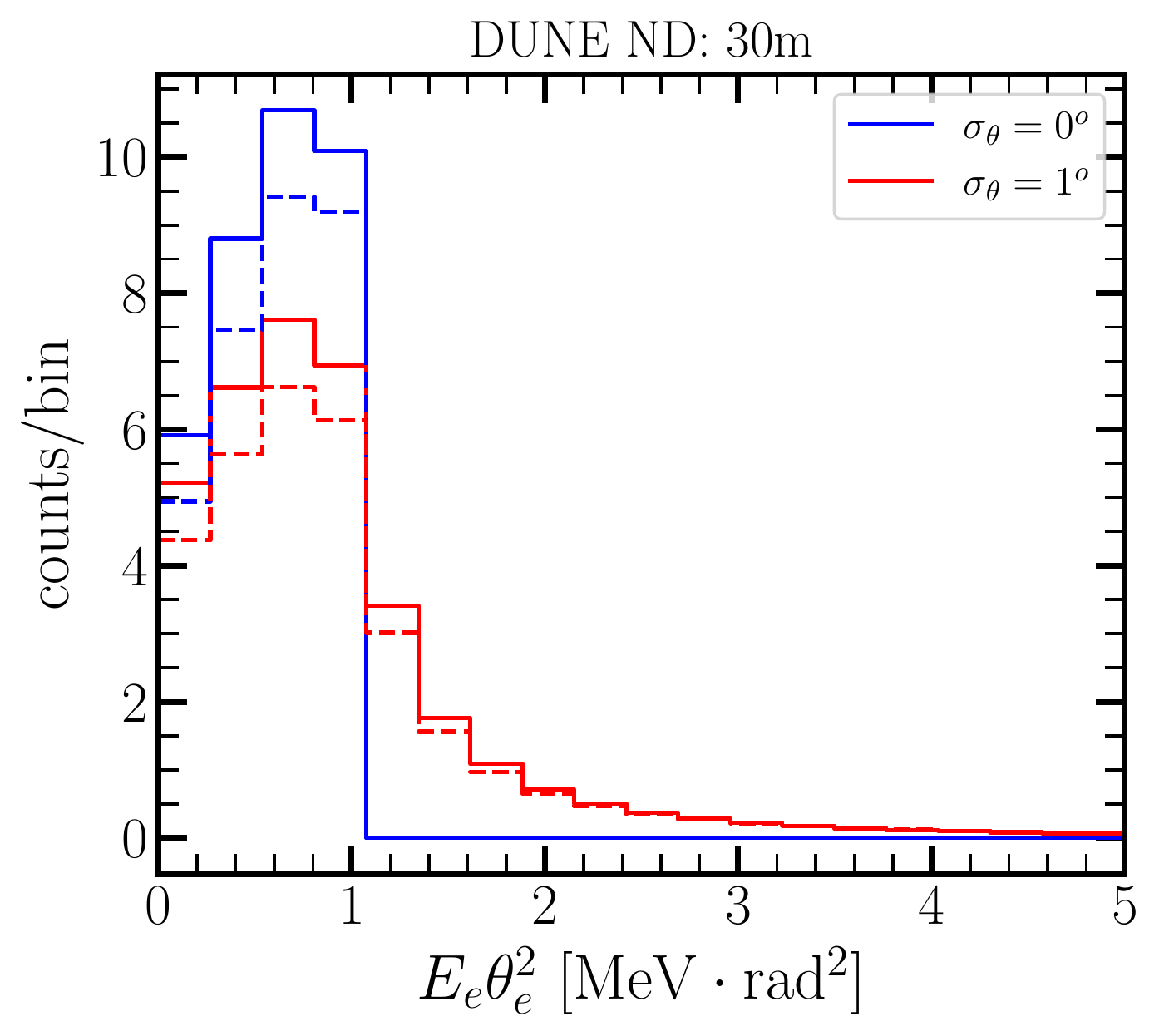}
\caption{Expected event spectra in the SM assuming 1 yr (neutrino) + 1 yr (antineutrino) mode at the DUNE-ND. Blue (red) spectra assume  perfect ($\sigma_\theta =1^o$) angular resolution, while solid (dashed) spectra correspond to neutrino (antineutrino) mode.}
\label{fig:SMevents}
\end{figure}

\begin{figure}[ht]
\centering
\includegraphics[width = 0.49 \textwidth]{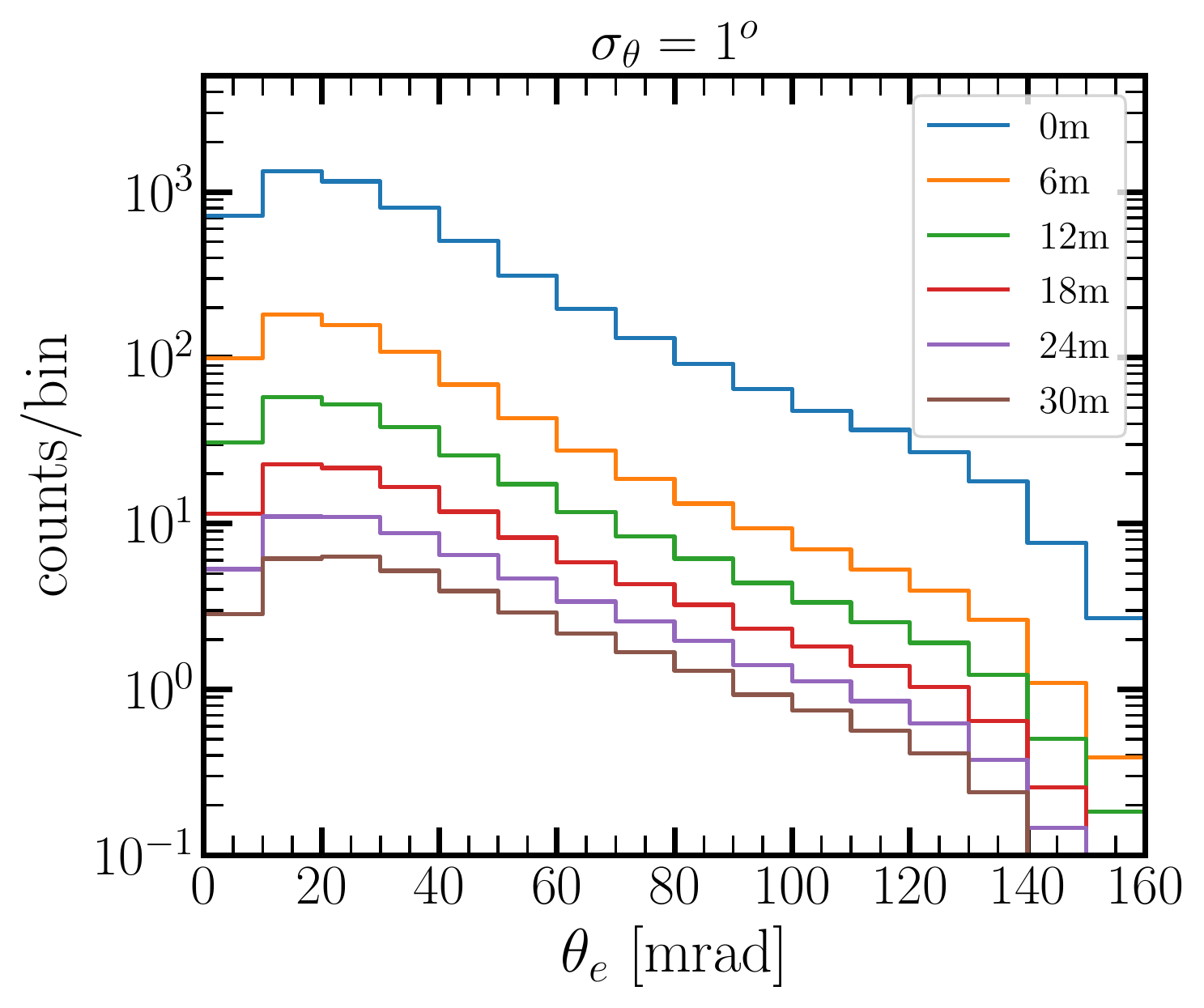}
\includegraphics[width = 0.49 \textwidth]{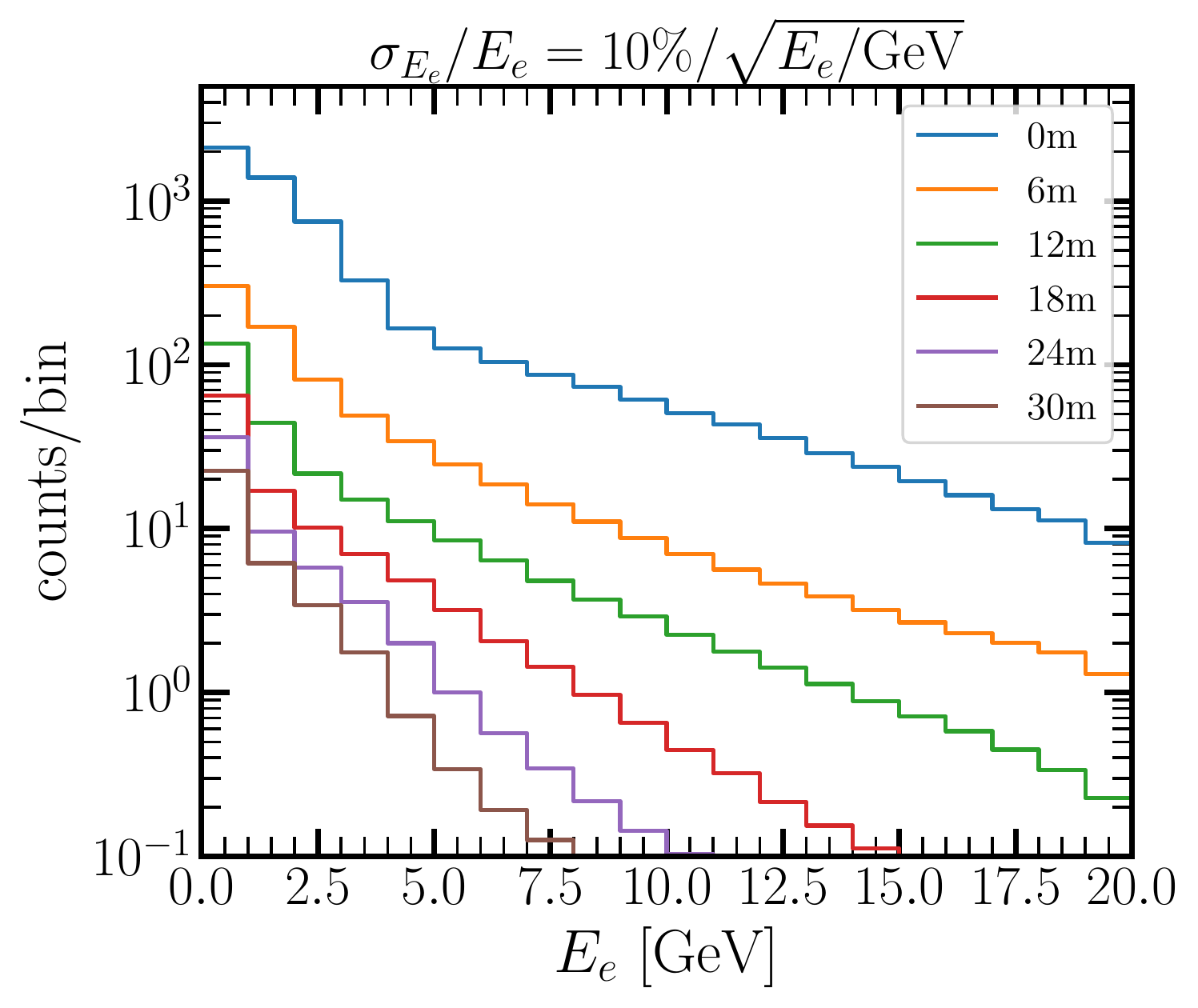}
\caption{Expected event spectra in the SM assuming 1 yr in neutrino mode at the DUNE-ND. The left (right) panel illustrates the rates in terms of the reconstructed scattering angle (electron energy).}
\label{fig:SMevents_recoil_theta}
\end{figure}

Assuming purely SM interactions, the expected event spectra at DUNE-ND corresponding to the different On-Axis and Off-Axis locations are depicted as a function of the true $(E_e \theta_e^2)^t$ (blue) and reconstructed $(E_e \theta_e^2)^\mathrm{reco}$ (red) in Fig.~\ref{fig:SMevents}. The calculation assumes 1 yr in neutrino mode (solid lines) and 1 yr in antineutrino mode (dashed lines), while for each mode the detector is assumed to be located half time On-Axis with the rest time shared between the various Off-Axis locations. As can be seen from the plot, the effect of angular smearing is rather significant allowing the calculated event spectra to extend far beyond the physical cutoff $E_e \theta_e^2 < 2 m_e$, as expected when resolution effects are ignored.\footnote{We have verified that our calculated event spectra are in excellent agreement with Ref.~\cite{deGouvea:2019wav}.} From the calculated spectra, it can be deduced that the expected signal will be dominated by the On-Axis induced-events which amounts to 82\% (81\%) of the total signal for the case of neutrino (antineutrino) mode.   Unless otherwise mentioned, in the remainder of the paper the present calculations will always assume reconstruction only for the case of the scattering angle, while the superscript ``reco" is dropped for convenience.
For completeness, we also show the expected \eves event rates projected in the reconstructed scattering angle and reconstructed electron energy space in the left and right panel of Fig.~\ref{fig:SMevents_recoil_theta}, respectively. We find that most of the electrons will scatter within a forward cone with angle 10--20 mrad with respect to the incident neutrino beam, while their angular distribution will not extend beyond 150 mrad. Regarding the electron energy distribution, most of the scattered electron population will have $E_e \leq 5$~GeV, while for energy $E_e \geq 10$~GeV the expected \eves signal is practically zero. A comparison of the $E_e \theta_e^2$, $\theta_e$ and $E_e$ distributions between the different locations is given in the Appendix~\ref{appendix1}.

\begin{figure}[ht]
\centering
\includegraphics[width = 0.32 \textwidth]{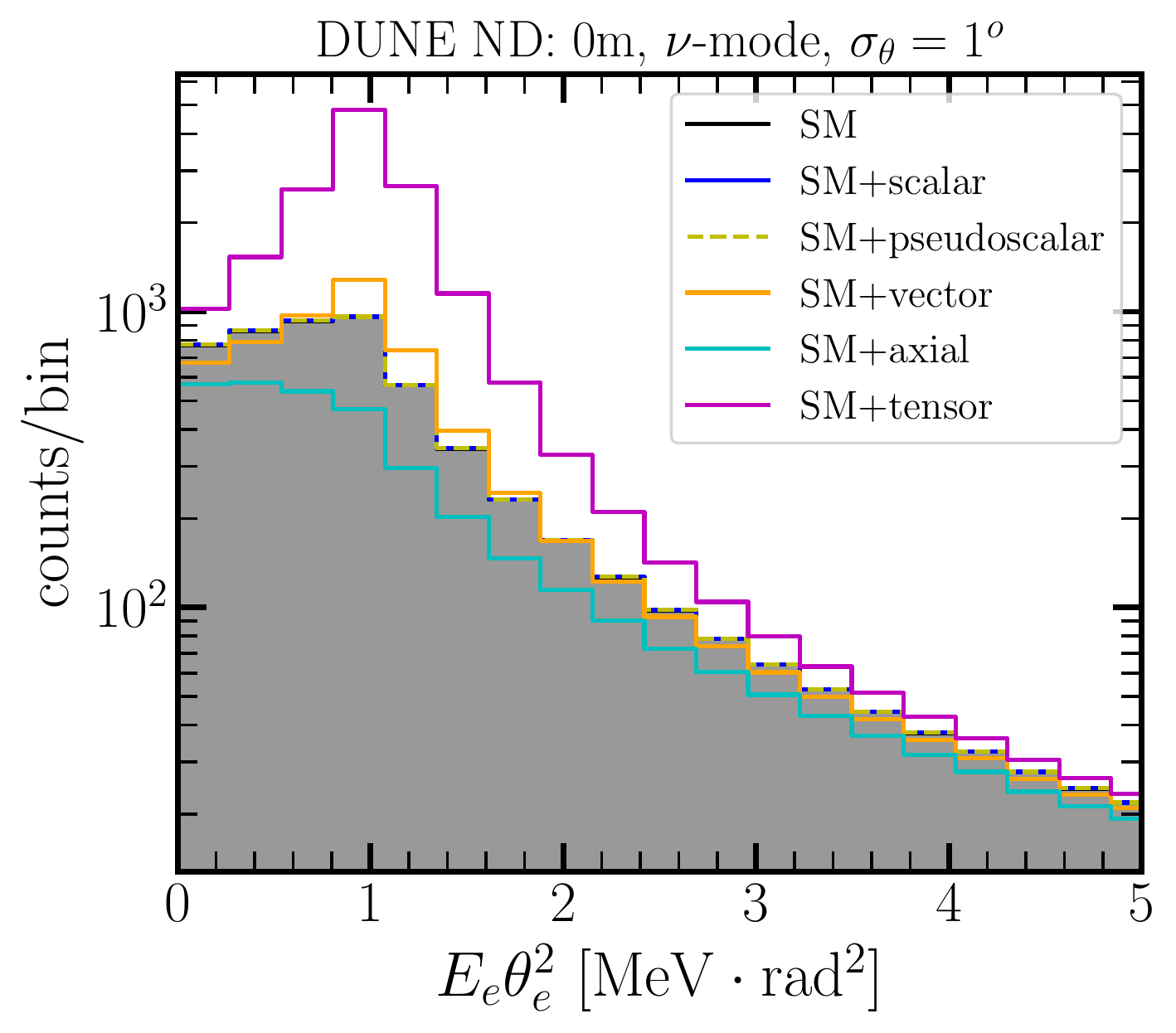}
\includegraphics[width = 0.32 \textwidth]{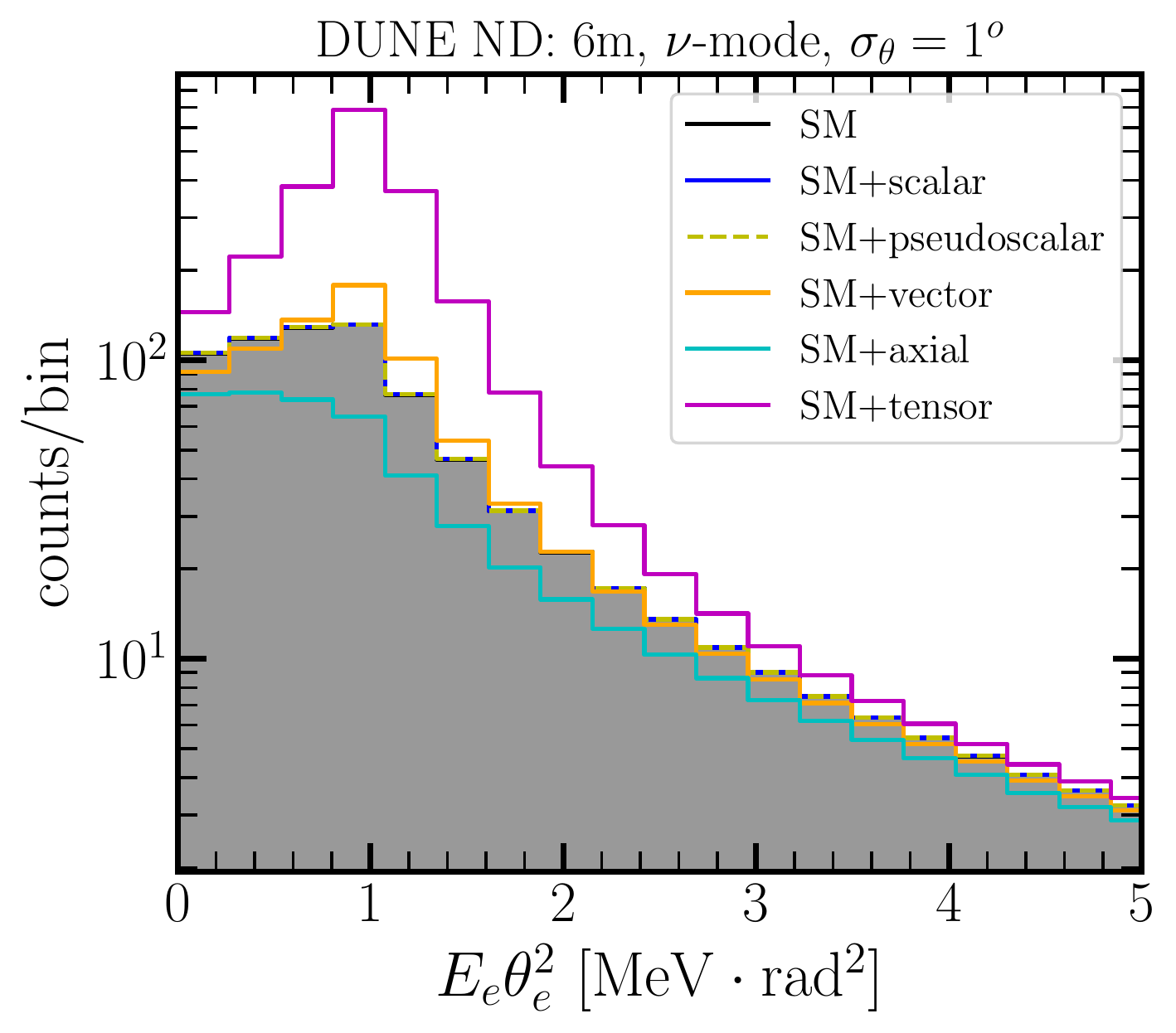}
\includegraphics[width = 0.32 \textwidth]{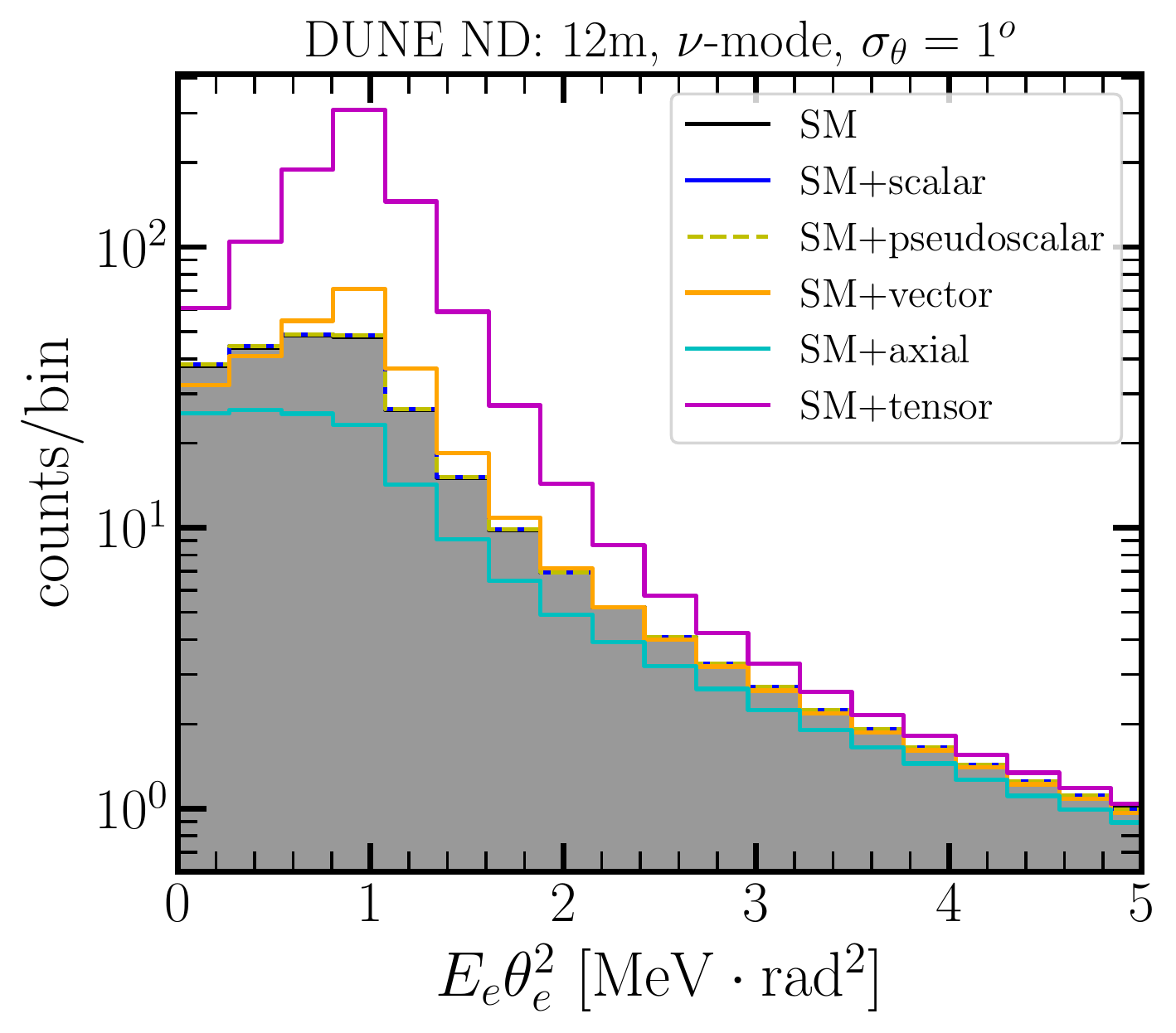}
\includegraphics[width = 0.32 \textwidth]{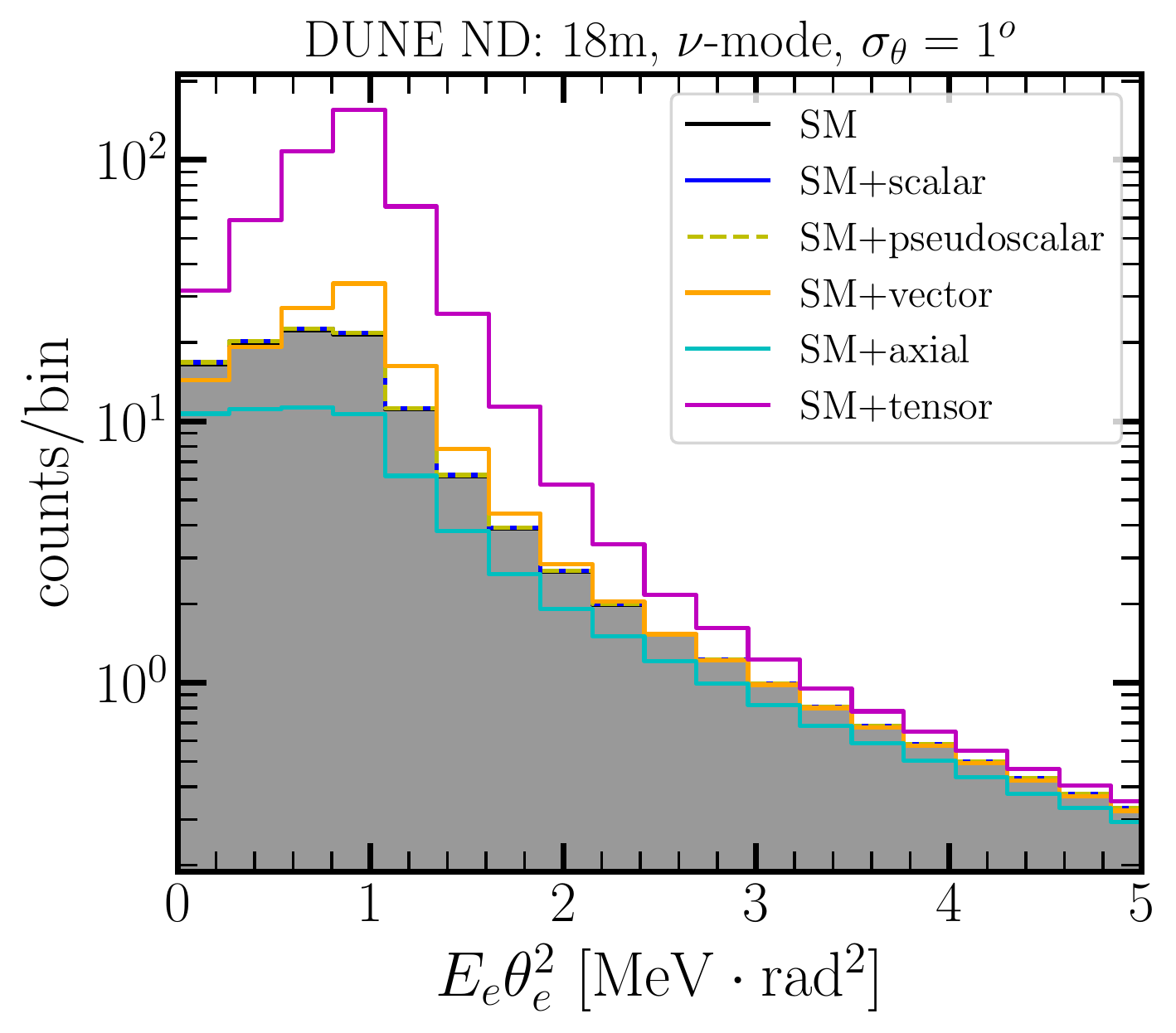}
\includegraphics[width = 0.32 \textwidth]{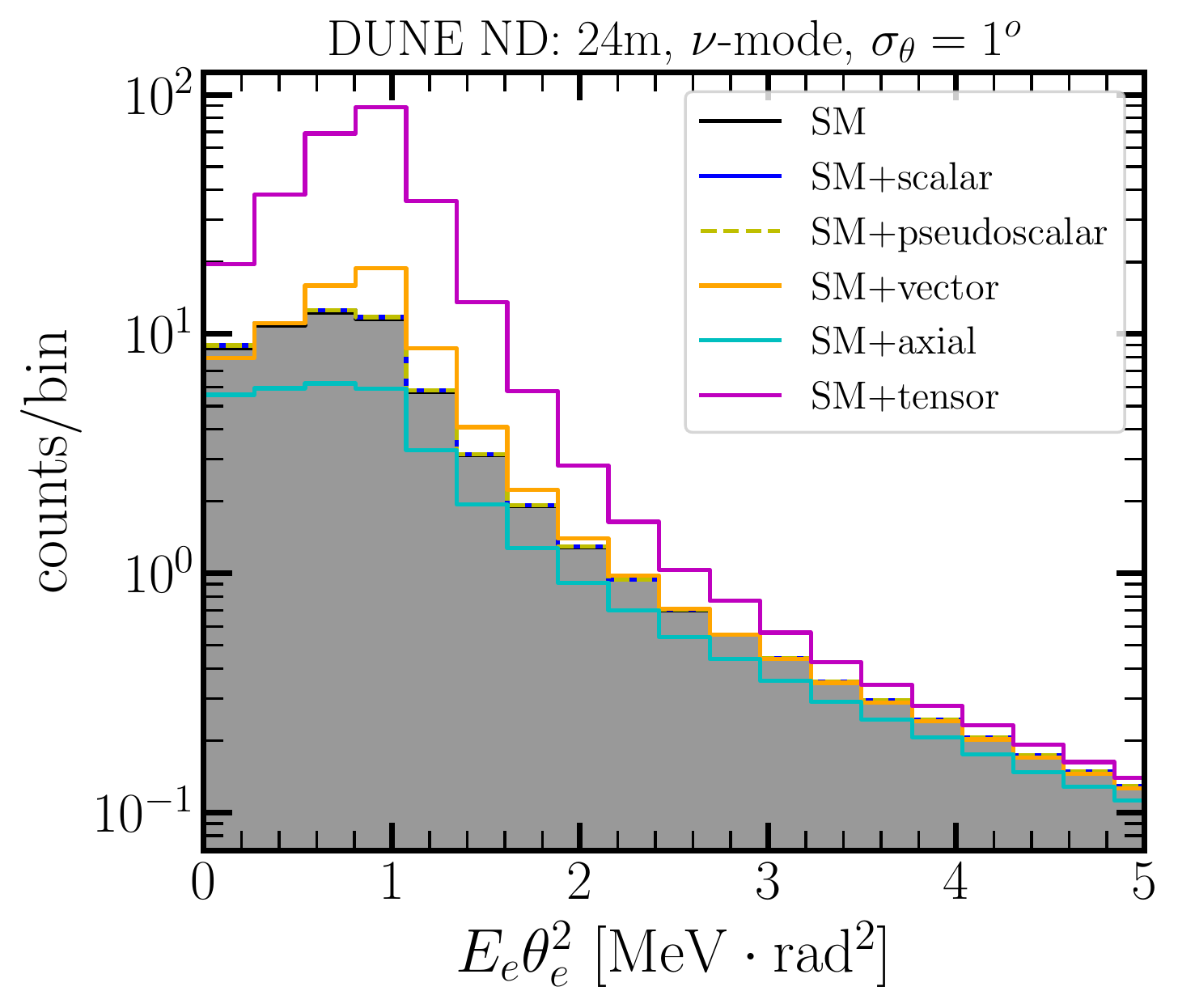}
\includegraphics[width = 0.32 \textwidth]{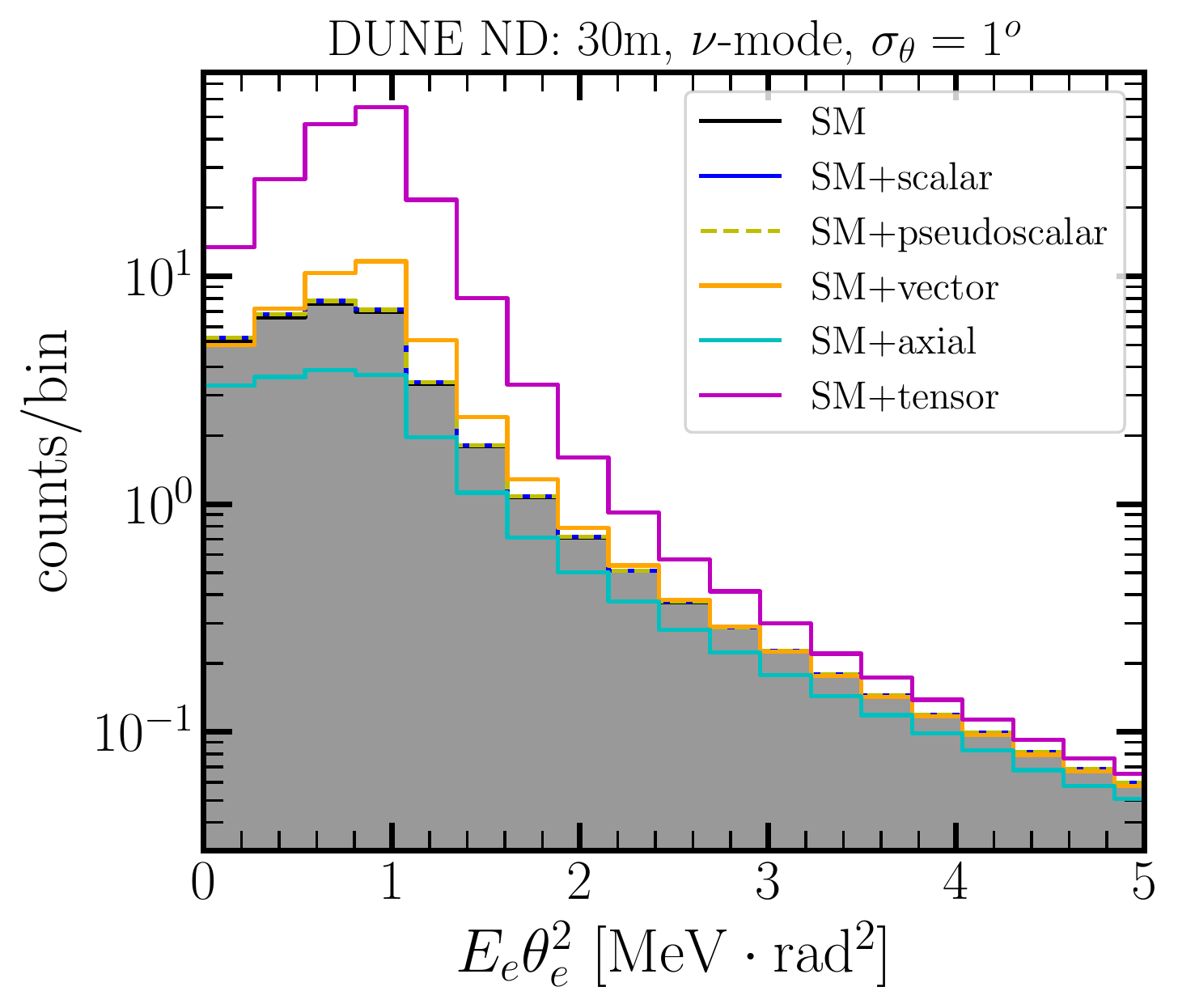}
\caption{Expected event spectra  at the different On-Axis and Off-Axis DUNE-ND locations in terms of $E_e \theta_e^2$ for the different interactions $X=S,P,V,A,T$ assuming 1 yr in neutrino mode and angular resolution with $\sigma_\theta =1^o$. The NGI spectra are calculated  for the benchmark parameters $g_X=5.7\cdot 10^{-5}$ and $m_X=10$~MeV.}
\label{fig:events_SPVAT}
\end{figure}

\begin{figure}[ht]
\includegraphics[width = 0.49 \textwidth]{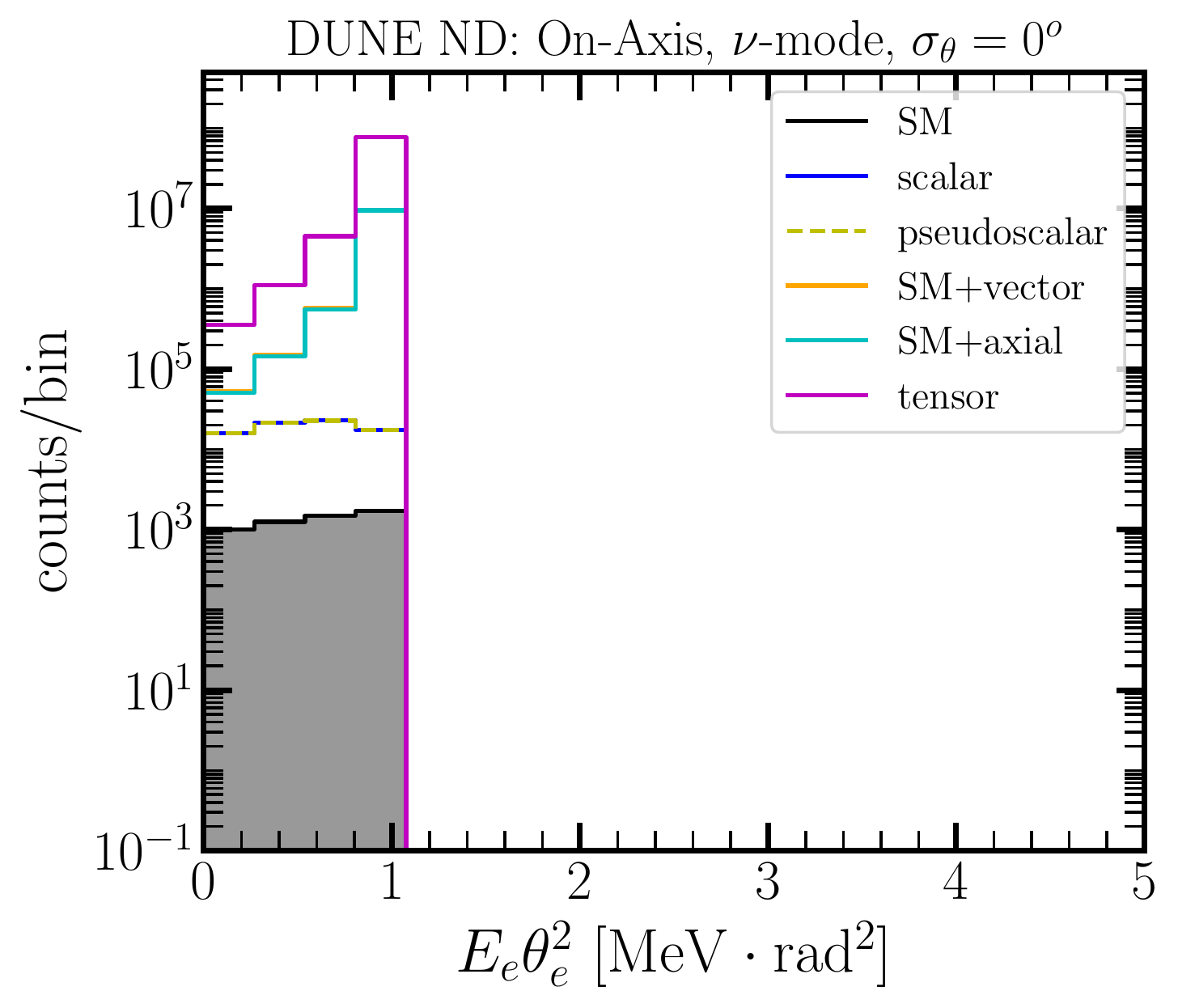}
\includegraphics[width = 0.49 \textwidth]{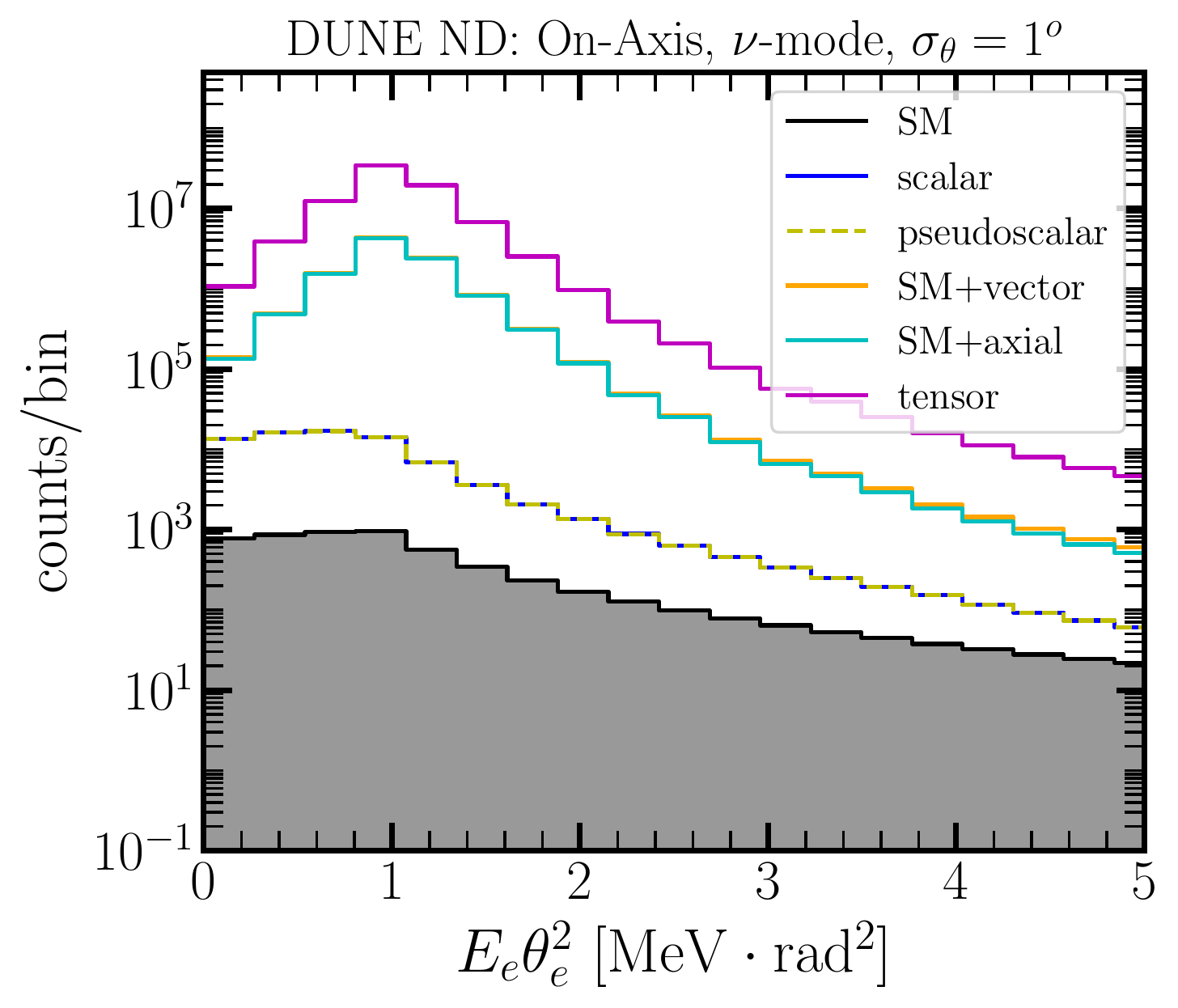}
\caption{Expected On-Axis event spectra in terms of $E_e \theta_e^2$ for the different interactions $X=S,P,V,A,T$ assuming 3.5 yr (neutrino) + 3.5 yr (antineutrino) mode at the DUNE-ND. The left (right) panel shows the results assuming perfect ($\sigma_\theta =1^o$) angular resolution.}
\label{fig:events_SPVAT2}
\end{figure}

Turning to beyond the SM scenarios, in Fig.~\ref{fig:events_SPVAT} we illustrate the expected NGI event spectra, calculated for the benchmark parameters $g_X=5.7\cdot 10^{-5}$ and $m_X=10$~MeV. The pure SM spectra are superimposed for comparison. It is also interesting to notice the effect of destructive interference for the case of the axial interaction. If present, NGIs may induce spectral features which makes them particularly interesting. For instance, the shapes of the scalar and pseudoscalar spectra are characteristically different compared to those of the rest interactions, as shown in the left and right panels of Fig.~\ref{fig:events_SPVAT2} assuming $g_X=4.5\cdot 10^{-4}$ and $m_X=150$~keV. Interestingly--and unlike the previous case--for the latter benchmark point a constructive  (destructive) interference is found for the vector  (axial vector) case.

\subsection{Sensitivity analysis at DUNE-ND}
Having developed the necessary machinery for accurately simulating the expected event spectra we proceed by performing a sensitivity analysis of the physics scenarios in question with the goal to explore the attainable sensitivities at DUNE-ND.
To this purpose,  we rely on the following $\chi^2$ function fitting simultaneously On-Axis and Off-Axis spectra and using shape and normalization information
\begin{equation}
    \chi^2 = 2 \sum_{k = \nu/\bar{\nu}} \, \sum_{j= \mathrm{loc}} \, \sum_{i=1}^{20} \left[N^{ijk}_\text{exp} - N^{ijk}_\text{obs}  + N^{ijk}_\text{obs} \, \log{\frac{N^{ijk}_\text{obs}}{N^{ijk}_\text{exp} } } \right] + \left( \frac{\alpha_1}{\sigma_{\alpha_1}}\right)^2 + \left( \frac{\alpha_2}{\sigma_{\alpha_2}}\right)^2 \, ,
\end{equation}
where the index $i$ runs over the reconstructed $E_e \theta_e^2$ bins for which we consider 20 evenly spaced values in the range $[0, 10~m_e]$, while  $j$ runs over the different On-Axis and Off-Axis locations of the ND i.e. $\{0,6,12,18,24,30\}~\mathrm{m}$, and finally $k$ accounts for neutrino and antineutrino mode. Here, the expected number of events are taken to be $N_\text{exp} = N_\text{SM} \cdot (1+ \alpha_1) + N_\text{bkg} \cdot (1+\alpha_2)$ with the background events being the sum of CCQE and missidentified $\pi^0$ events as $N_\text{bkg} = N_{\pi^0}^\text{missID} + N_\text{CCQE}$. For the observed events we consider the sum of SM and new physics \eves events namely $N_\text{new}$ that we add to the background events. Note, that  $N_\text{new}=N_\text{SM} + N_{X}(g_X, m_X)$  for $X=S,P,T$ since there is no interference between the interaction $X$ and the SM, while for $X=V,A$, due to interference with the SM one has  $N_\text{new}=N_{X}(g_X, m_X)$, i.e the SM contribution is included in $N_X$. Similarly, for the case of $E_6$ one has $N_\text{new}=N_{E_6}(m_{Z^\prime}, \cos \beta)$ while for left-right symmetry $N_\text{new}=N_{\mathrm{LR}}(m_{Z^\prime})$. We furthermore consider two nuisance parameters $\alpha_1$ and $\alpha_2$ with $\sigma_{\alpha_1}=5\%$ and $\sigma_{\alpha_2}= 10\%$ to account for the normalization uncertainties of the DUNE neutrino flux and background, respectively.

At this point we wish to devote a separate paragraph in order to discuss our current assumptions regarding the background event rates. Since we have not available background events corresponding to the various Off-Axis locations we simply rescale the On-Axis background rates by the exposure time, achieving a conservative estimate. While this approximation neglects the shape of the background events corresponding to Off-Axis locations, we however expect this to have a rather negligible impact on our sensitivities for the following reasons. First because our sensitivities are dominated by the On-Axis events, and second because we expect that shape uncertainties are indirectly accounted for in the rather large background normalization uncertainty.

\subsection{Other experiments}

\subsubsection{COHERENT}
\label{sec:coherent}   

The COHERENT collaboration has measured \cevns events on a 14.57~kg CsI~\cite{COHERENT:2017ipa,COHERENT:2021xmm} detector and on a 24~kg liquid argon (LAr)~\cite{COHERENT:2020iec} detector, both located at the Spallation Neutron Source (SNS).
At the SNS muon neutrinos are produced from pion decay at rest ($\pi$-DAR) $\pi^+ \to \mu^+ + 
\nu_\mu$ which are prompt with the beam, while delayed $\nu_e$ and $\bar{\nu}_\mu$ are generated from the subsequent muon decay $\mu^+ \to e^+ + \nu_e + \bar{\nu}_\mu$. We employ the Michel spectrum which adequately describes the neutrino distributions~\cite{Louis:2009zza}.
We simulate the event spectra for the CsI\footnote{For CsI, we consider the full data reported in 2021~\cite{COHERENT:2021xmm}.} and LAr COHERENT detectors taking into account the detector specifications, systematic uncertainties and backgrounds\footnote{Namely, steady state backgrounds and beam related neutrons.} and we express our reconstructed signal using the time and energy binning reported by the corresponding data release (see Refs.~\cite{COHERENT:2021xmm, COHERENT:2020iec}). In particular, for the CsI measurement we consider 11 time bins in the range $[0,6]~\mathrm{\mu s}$ and 9 energy bins expressed in photoelectrons (PE) in the range $[0,60]$~PE. Similarly for the case of LAr we consider 10 time bins in the range $[0,5]~\mathrm{\mu s}$ and 12 energy bins expressed in units of electron equivalent energy in the range $[0,120]~\mathrm{keV_{ee}}$.
An as realistic as possible sensitivity analysis is performed following the analysis strategy of Ref.~\cite{DeRomeri:2022twg}. Let us finally note that even though our results regarding COHERENT are mainly driven by the CsI data, we nonetheless perform a combined analysis of CsI+LAr data.

\subsubsection{Dresden-II}
\label{sec:dresdenII}
For the first time, a suggestive evidence of \cevns was recently announced by the Dresden-II Collaboration using reactor antineutrinos ~\cite{Colaresi:2022obx}. The experiment used a 3 kg germanium detector exposed to reactor antineutrinos emitted from the Dresden-II boiling water reactor, and collected data for an ON/OFF period corresponding to 96.4/25 days. Due to its close proximity to the reactor core the experiment is dominated by epithermal neutron backgrounds and electron capture peaks in $^{71}$Ge. During the beam ON period a very strong preference over the background-only hypothesis was found in the data that is consistent with a CE$\nu$NS-induced signal in the low-energy measured spectrum, while the data collected during the beam OFF period are consistent with the null hypothesis. We consider the antineutrino spectrum from  Ref.~\cite{CONNIE:2019xid} taking into account all the relevant fissible  isotopes, i.e., $^{235}$U, $^{238}$U, $^{239}$Pu, $^{242}$Pu  and $^{238}\mathrm{U}(n, \gamma)$ as well as the natural abundances of Ge, as done in Ref.~\cite{Majumdar:2022nby}. To simulate the background, we reproduce the background model parametrized in terms of seven free parameters as detailed in the supplemental material of Ref.~\cite{Colaresi:2022obx}. Our simulated signal is then expressed in terms or reconstructed electron equivalent energy using 130 bins evenly distributed over the range $[0.2,1.5]~\mathrm{keV_{ee}}$.  Finally,  our statistical analysis proceeds through a simultaneous fit of background model parameters and new physics model parameters following Ref.~\cite{Majumdar:2022nby}.

\subsubsection{TEXONO} 
\label{sec:texono}

In the present work we also consider existing data of measured 
\eves events reported by the TEXONO collaboration~\cite{TEXONO:2009knm}. The experiment observed \eves exploiting reactor antineutrinos emerged from the Kuo-Sheng Nuclear reactor using a 187~kg CsI(Tl) detector. 
Unlike the Dresden-II case, for the case od TEXONO we adop the reactor antineutrino energy distribution  from~\cite{Mention:2011rk} for  $E_\nu> 2~\mathrm{MeV}$, while for $E_\nu < 2~\mathrm{MeV}$ we consider the theoretical estimations from Ref.~\cite{Kopeikin:1997ve} and we finally assume an overall normalization of  $6.4 \times 10^{12}~\mathrm{cm^{-2} s^{-1}}$.
Our simulated signal is expressed in units of $\mathrm{events/(kg\cdot day \cdot MeV)}$ for which we consider 10 bins in the range $[3,8]$~MeV following Ref.~\cite{TEXONO:2009knm}, while our statistical analysis is based on the assumptions of Ref.~\cite{Miranda:2021kre}.

\section{Results and discussion}
\label{sec:results}

\begin{figure}[t]
\centering
\includegraphics[width = 0.49 \textwidth]{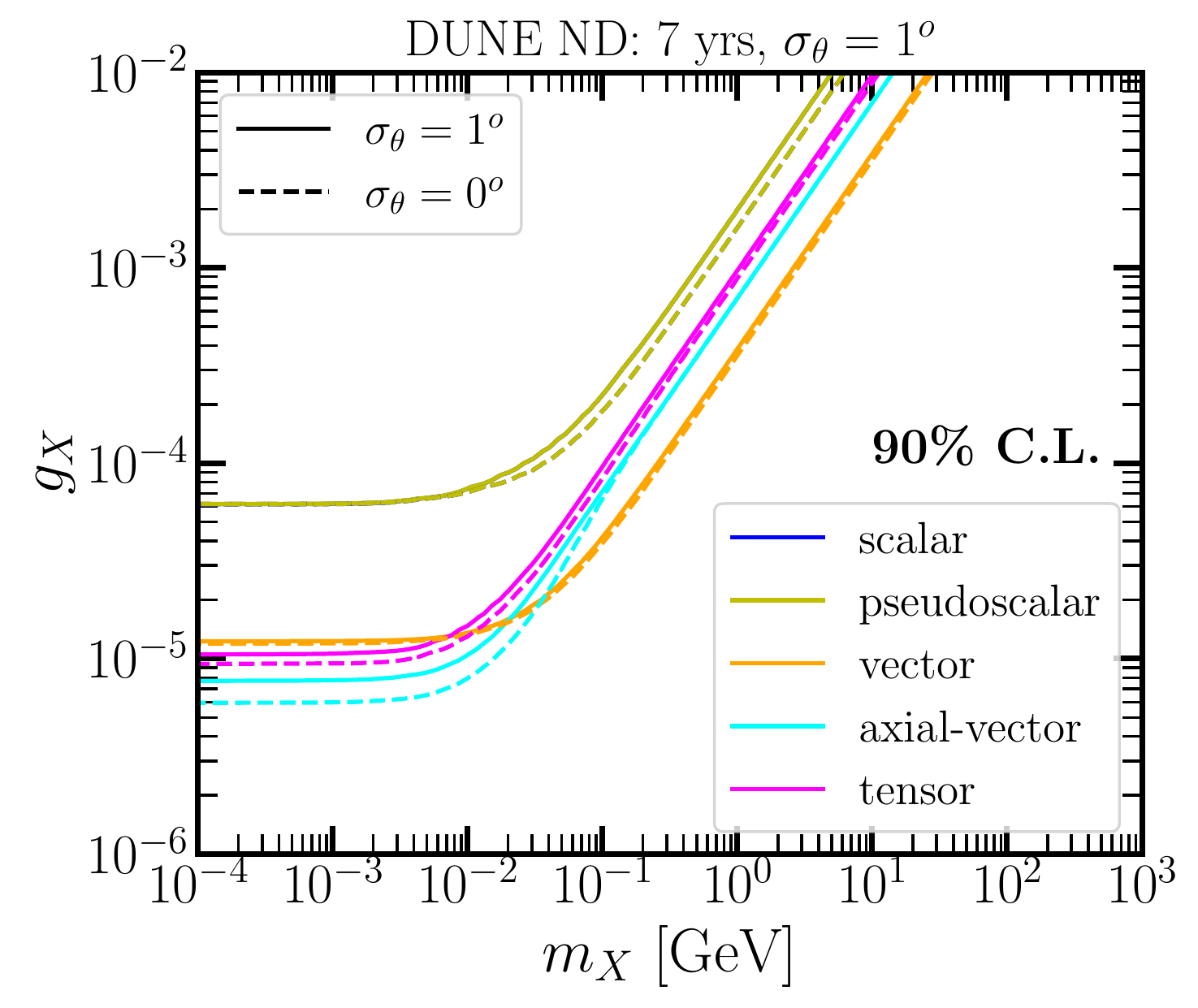}
\caption{\textbf{Upper-Left:} Sensitivity of DUNE-ND on the different interactions $X=S,P,V,A,T$ assuming the 3.5 yr (neutrino) + 3.5 yr (antineutrino) mode. The results are illustrated at 90\% C.L. for ideal (dashed curves) and $\sigma_\theta = 1^o$ resolution (solid curves). The scalar and pseudoscalar contours are identical (see the text).}
\label{fig:SPVAT_contours}
\end{figure}

We now turn to our statistical analysis, by first focusing on the NGI sensitivities. Figure~\ref{fig:SPVAT_contours} illustrates the excluded regions at 90\% C.L. after 7 yrs of neutrino data collection at DUNE-ND, for the various interactions $X$.  We have verified that the sensitivity remains essentially unchanged when a total running time of 5 yr (neutrino mode) + 5 yr (antineutrino mode) is assumed. In the depicted contours, three regions can be  identified depending on the momentum transfer $q=\sqrt{2 m_e T_e}$ :  (i) $m_X \ll q$ corresponding to light mediators  giving a line
parallel to the $m_X$ axis ,
(ii) $m_X \sim q$ which gives the turning points and
(iii) $m_X \gg q$ which corresponds to the heavy mediator case giving the part which rises with $m_X$. Contrary to case (i), the existence of a heavy NGI mediator does not modify the shape of the  spectra, and corresponds to a NGI scenario with effective couplings. To highlight the impact of angular resolution, the results are demonstrated for the case of ideal resolution (dashed lines) as well as for $\sigma_\theta = 1^o$ (solid lines), from where a slight reduction of the sensitivity reach  becomes evident. The depicted contours allow for a relative comparison of the expected sensitivity on the various NGIs from where it can be deduced that the least (most) constrained is the   scalar/pseudoscalar (axial vector) case.  This is due to the fact that the   scalar/pseudoscalar cross section is suppressed by a factor   $T_e^2/E_\nu^2$ compared to the leading terms of either vector, axial vector or tensor cross sections.   As explained above, the scalar and pseudoscalar sensitivities are practically identical. Furthermore, at low mediator masses the interaction channels $X=V,T$ have comparable sensitivity while the constraint for $X=A$ is exceeding that of $X=T$. To clarify this behavior, let us highlight that while the tensor interaction cross section is larger by a factor 8 compared to the purely NGI vector or axial vector ones, there is an improvement in the sensitivity of $X=A,V$ cases  due to the existence of interference with the SM.  Before we proceed--and in connection to the background-related discussion made above--let us stress that we have furthermore checked that increasing the background by a factor two has no visible impact on the results.

\begin{figure}[ht]
\centering
\includegraphics[width = 0.49 \textwidth]{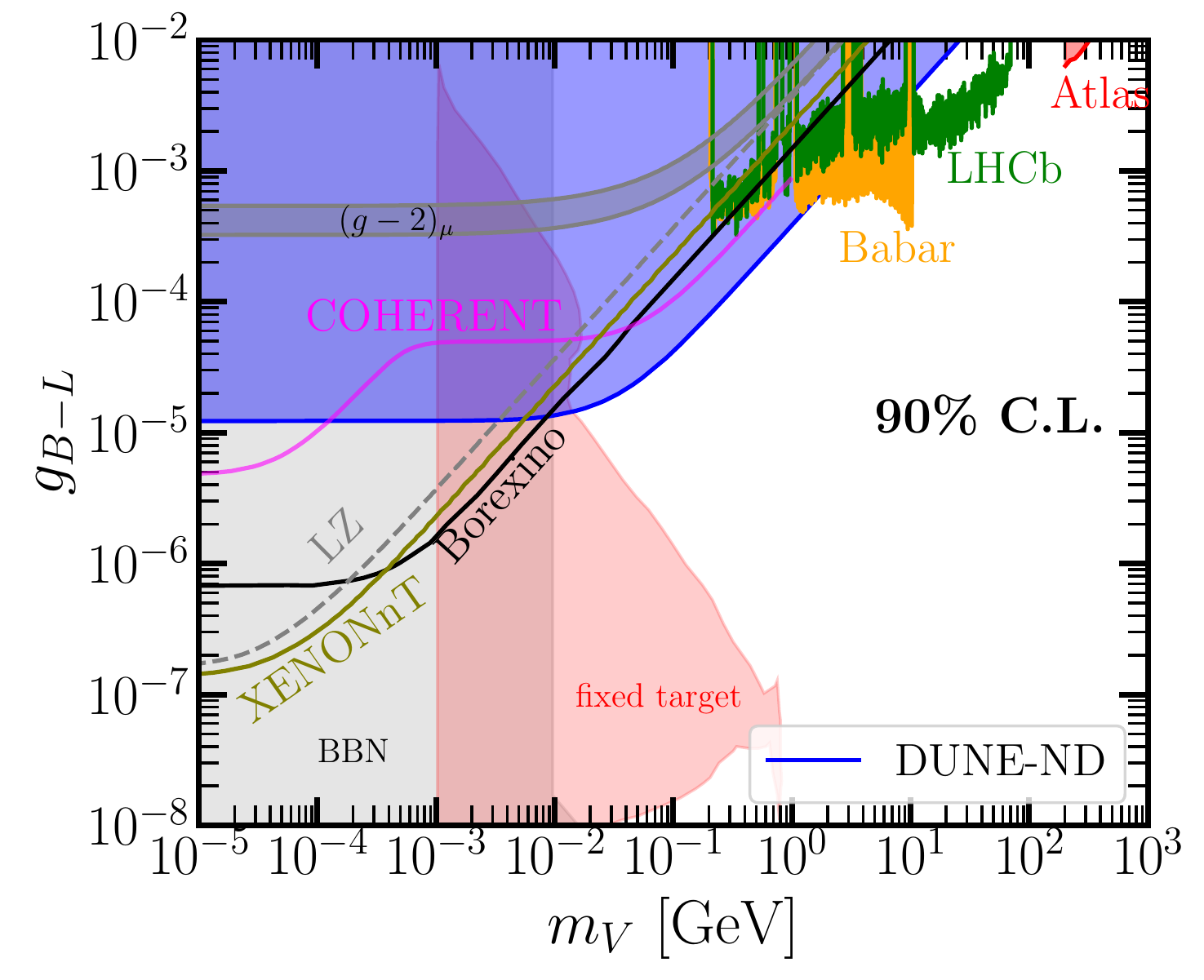}
\includegraphics[width = 0.49 \textwidth]{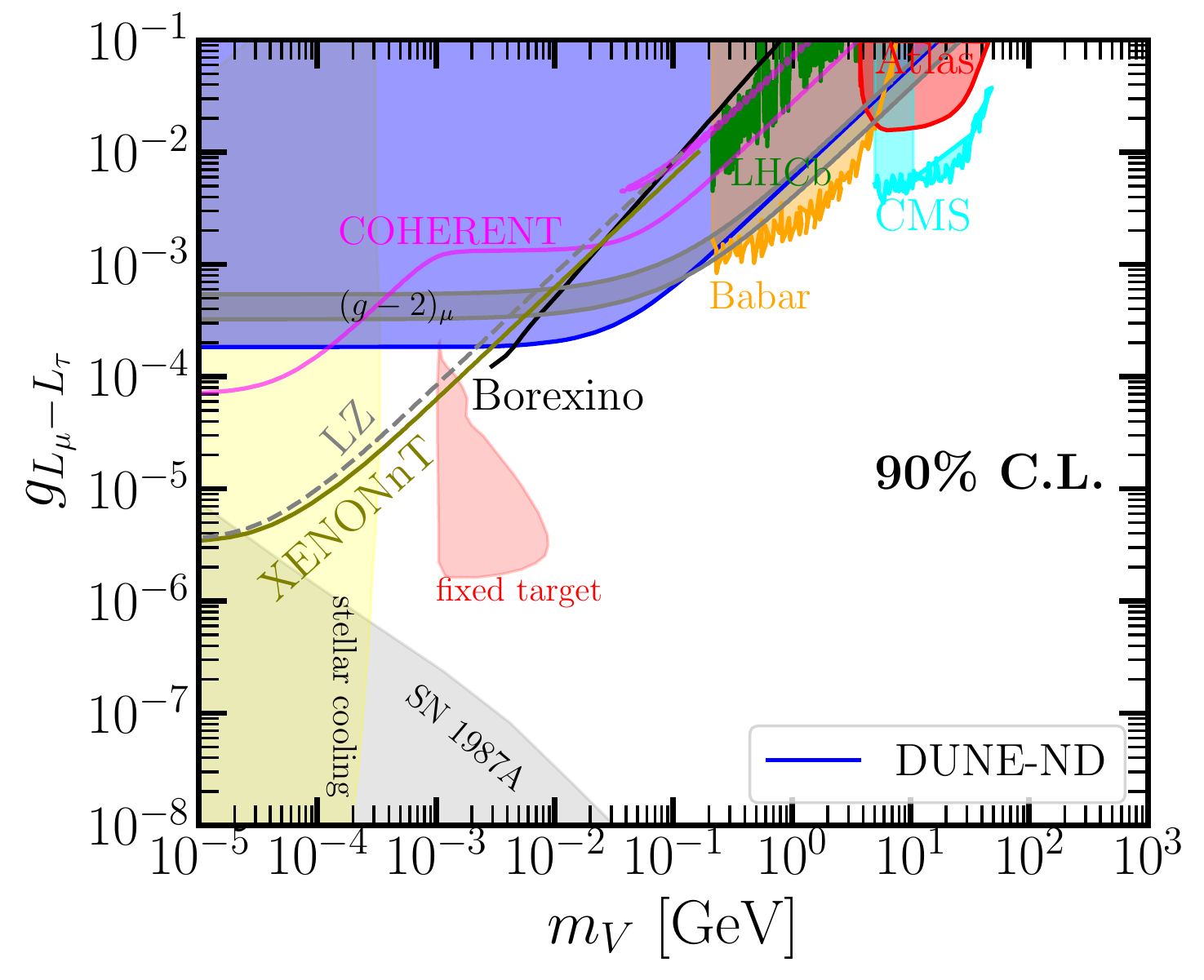}
\caption{Comparison of the DUNE-ND sensitivity with existing constraints (see the text for details). The results are presented at 90\% C.L. for the case of universal vector $B-L$ (left) and $L_\mu - L_\tau$ (right) interactions.}
\label{fig:compare}
\end{figure}

\begin{figure}[ht]
\centering
\includegraphics[width = 0.49 \textwidth]{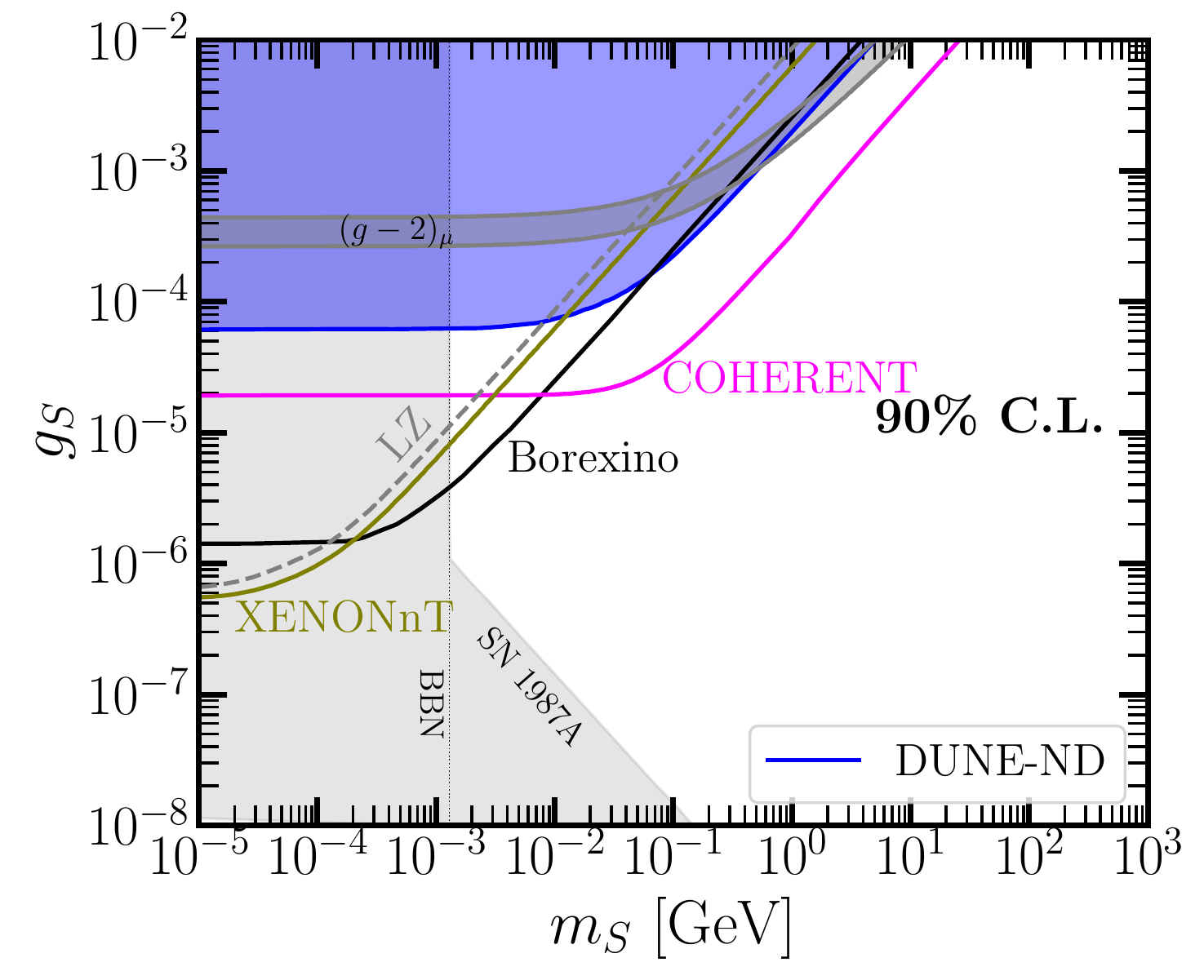}
\includegraphics[width = 0.49 \textwidth]{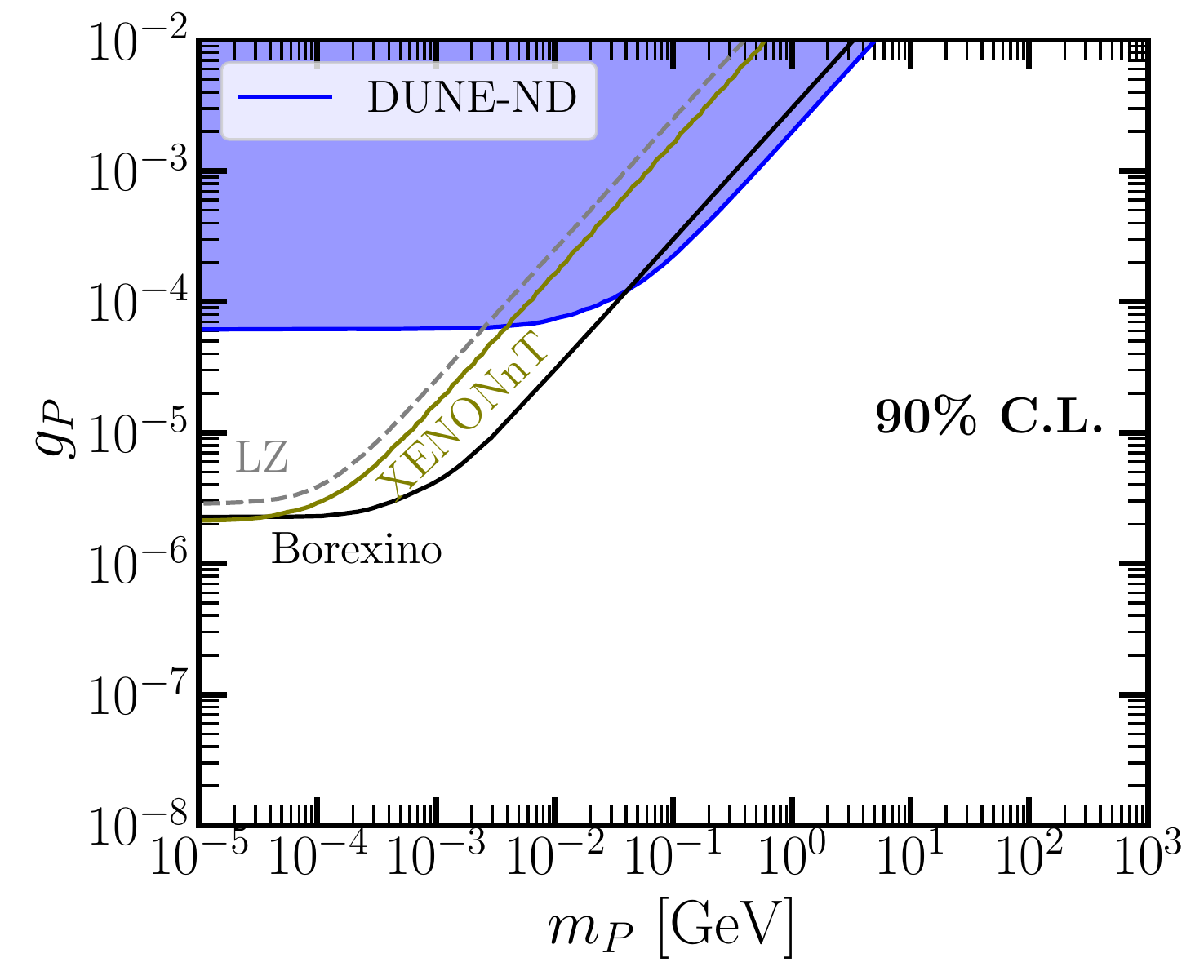}

\includegraphics[width = 0.49 \textwidth]{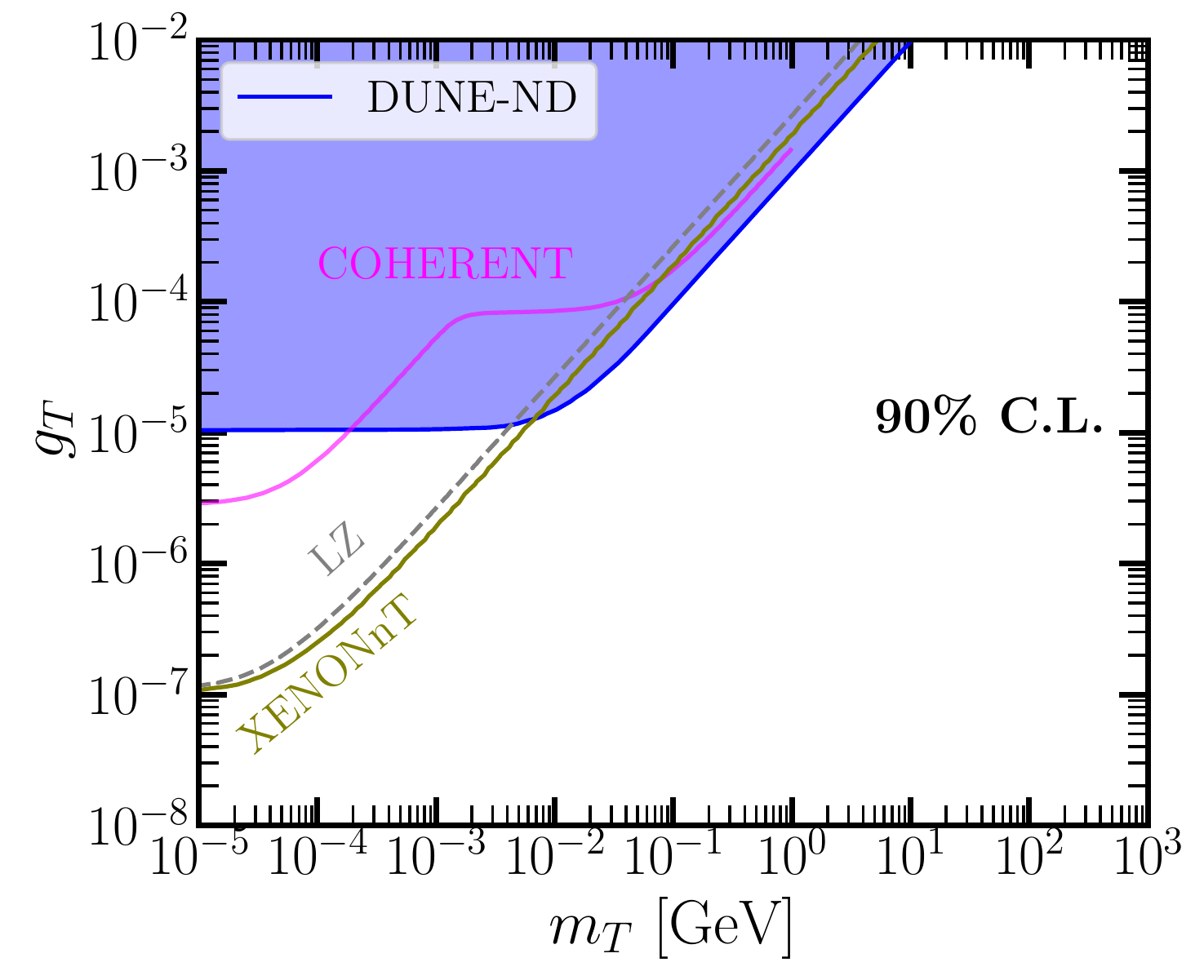}
\includegraphics[width = 0.49 \textwidth]{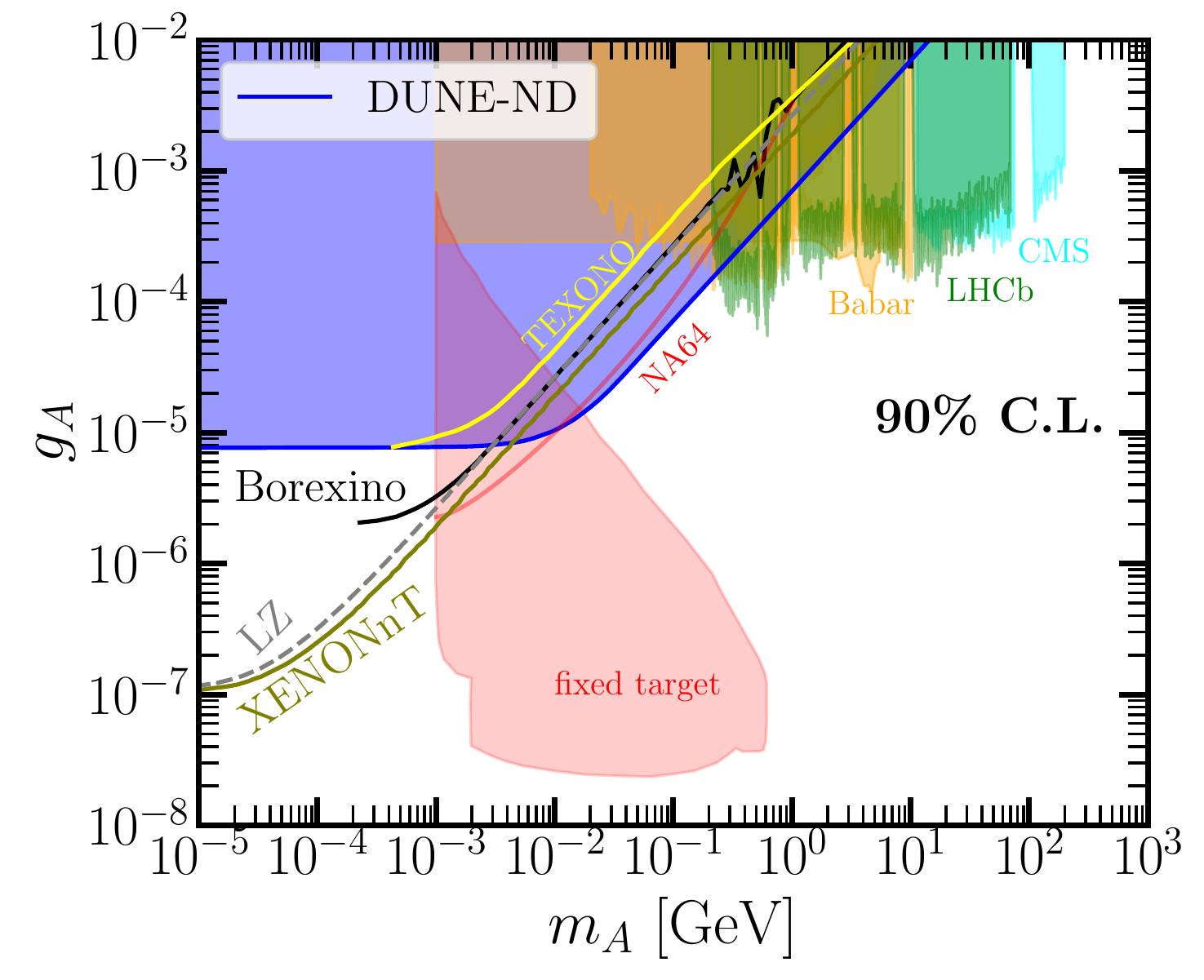}
\caption{ Comparison of the DUNE-ND sensitivity with existing constraints (see the text for details). The results are presented at 90\% C.L. for the case of scalar (top-left), pseudoscalar (top-right), tensor (bottom-left) and axial vector (bottom-right) interactions.}
\label{fig:compare2}
\end{figure}

It is interesting to compare the attainable DUNE-ND sensitivities obtained in this work with existing constraints coming from further experimental probes. As pointed out in Ref.~\cite{AristizabalSierra:2020edu}, for the case of \eves the general vector case discussed up to now is identical to the anomaly-free vector $B-L$ scenario, thus in order to compare our present results we choose the latter. Then, in Fig.~\ref{fig:compare} we show the limits for the case of vector $B-L$ (left), $U(1)_{L_\mu - L_\tau}$  (right) gauge symmetric models  by superimposing existing limits from various experimental probes.  Specifically, we compare the DUNE-ND with low-energy neutrino experiments such as those looking for \cevns like COHERENT~\cite{COHERENT:2020iec, COHERENT:2021xmm} and solar neutrinos e.g. Borexino~\cite{Borexino:2017rsf}. The COHERENT (combined CsI+LAr data) sensitivities for the vector $B-L$  case are taken from the analysis of Ref.~\cite{DeRomeri:2022twg}~\footnote{Let us note that Ref.~\cite{DeRomeri:2022twg} obtained constraints for the general vector case, which here is recasted to the $B-L$ case as follows: for $m_V > 300~\mathrm{keV}$ which corresponds to the CE$\nu$NS-induced region, the $B-L$ COHERENT limit is roughly obtained from the general vector case reported in~\cite{DeRomeri:2022twg} through the shift $g_{B-L} \approx 3 g_V$, while for $m_V < 300~\mathrm{keV}$ corresponding to the low-energy E$\nu$ES-induced region it holds  $g_{B-L} = g_V$. Finally the narrow region corresponding to destructive interference (see the left panels of Figs.~8 and 10 in Ref.~\cite{DeRomeri:2022twg}) is irrelevant for the $B-L$ case considered here.}, while for the $L_\mu - L_\tau$ scenario the sensitivities are extracted in the present work following the analysis of the latter study.  Similarly, the XENONnT and LZ constraints for the vector $B-L$ case are adapted from Ref.~\cite{A:2022acy}, while the  $L_\mu - L_\tau$ constraints are obtained for the first time in this work.  Note, that the vector bounds driven by \eves events taken from Refs.~\cite{DeRomeri:2022twg,A:2022acy} are appropriately adjusted according to $g_{B-L}\to g_{B-L}/\sqrt{2}$  to be consistent with the cross section definition used in the present work. \footnote{Projected sensitivities from futuristic measurements at direct dark matter detection experiments are reported in Ref.~\cite{Majumdar:2022nby}.} Turning to solar neutrino analyses, for the vector $B-L$ case the constraints are taken from Ref.~\cite{Coloma:2022umy} which performed a spectral analysis of Borexino Phase-II data, while the   $L_\mu - L_\tau$ limits are taken from the analysis of  Ref.~\cite{Amaral:2020tga,Gninenko:2020xys}. We furthermore show the corresponding limits obtained  from high energy collider data, recasted to the vector $B-L$  and $L_\mu - L_\tau$ cases using the \href{https://gitlab.com/philten/darkcast}{Darkcast} software package.  Specifically, we illustrate limits from  electron beam-dump\footnote{These include
E141~\cite{Riordan:1987aw,Bjorken:2009mm},
E137~\cite{Bjorken:1988as,Andreas:2012mt},
E774~\cite{Bross:1989mp},
KEK~\cite{Konaka:1986cb},
Orsay~\cite{Andreas:2012mt},
U70/$\nu$-CAL~I~\cite{Blumlein:2011mv,Blumlein:2013cua},
CHARM~\cite{CHARM:1985anb,Gninenko:2012eq},
NOMAD~\cite{NOMAD:2001eyx}, NA64~\cite{NA64:2016oww,NA64:2019auh},
A1~\cite{Merkel:2014avp} and APEX~\cite{APEX:2011dww}. Note that for the $L_\mu - L_\tau$ scenario only E141, E137 and KEK limits are relevant.}  experiments~\cite{Harnik:2012ni,Ilten:2018crw} as well as limits from ATLAS~\cite{Aaboud:2016cth} extracted from dielectron resonance measurements. Also shown are existing $B-L$ limits from  BaBar~\cite{Lees:2014xha,Lees:2017lec}, CMS~\cite{CMS:2019kiy} and LHCb~\cite{LHCb:2019vmc} Dark Photon analyses. Regarding $L_\mu - L_\tau$, we include limits from BaBar~\cite{BaBar:2016sci}, CMS~\cite{CMS:2018yxg}, and ATLAS~\cite{Altmannshofer:2016jzy,ATLAS:2014jlg} obtained from $4 \mu$ searches and LHCb~\cite{LHCb:2019vmc}.\footnote{These limits are obtained from $A^\prime \to \mu^+ \mu^-$ and recasted to the $L_\mu - L_\tau$ scenario.}  Astrophysical limits from Big Bang Nucleosynthesis (BBN)~\cite{Blinov:2019gcj,Suliga:2020jfa}, stellar cooling and SN 1987A~\cite{Croon:2020lrf} are also shown for comparison. Finally the limits on $(g-2)_\mu$ are obtained based on the Appendix B of Ref.~\cite{AtzoriCorona:2022moj}.  As a general remark, it should  be stressed that in all cases, DUNE-ND is expected to place competitive constraints to those derived from low-energy solar, dark matter direct detection and \cevns experiments and complementary to high energy collider searches~\cite{Das:2021esm}.

In Fig.~\ref{fig:compare2} we show a comparison of the sensitivity contours for the remaining scalar (top-left), pseudoscalar (top-right), tensor (bottom-left) and axial vector (bottom-right) interactions. For the scalar case astrophysical limits coming from BBN and SN 1987 are relevant, while for the tensor and pseudoscalar cases existing limits are coming from solar neutrinos (Borexino), direct dark matter detection  (LZ and XENONnT) and \cevns experiments (COHERENT). For the axial vector case relevant limits come from beam-dump experiments and collider searches which we reproduce  from Ref.~\cite{Baruch:2022esd} using \href{https://gitlab.com/philten/darkcast}{Darkcast}. Let us finally note that for the pseudoscalar and axial vector cases \cevns limits are not as competitive since they are suppressed by nuclear spin, and hence not shown here. 

A few comments are in order. The NGI cross section is enhanced significantly for low $T_e$, thus making the dark matter direct detection experiments favorable locations to probe the parameter space with very low masses $m_X$. Indeed, as can be seen from the plots, LZ and XENONnT dominate the limits for low mediator masses because of their very low-energy threshold detection capabilities. On the other hand, COHERENT dominates the limits in the parameter space for $m_X >10~\mathrm{MeV}$. However, it is important to clarify that CE$\nu$NS-based limits such as those coming from COHERENT are extracted from nuclear recoil measurements, and therefore they apply only under the assumption of universal couplings between the mediators $X=S,P,V,A,T$ and electrons $g_{e X}$ or quarks $g_{q X}$. Thus, in the general case the DUNE-ND by  exploiting the highly intense LBNF beam--which peaks in the ballpark of few GeV neutrino energy--has the prospect to place the most stringent limits for $m_X > 100~\mathrm{MeV}$ up to few GeV.
It is also interesting to notice that for the special case of $L_\mu - L_\tau$ symmetric model, DUNE-ND is expected to rule out the COHERENT limits with significant complementarity to LZ, XENONnT and Borexino. This is mainly because of the highly intense muon neutrino beams available at Fermilab. For the scalar  and pseudoscalar cases instead, the projected DUNE-ND sensitivities will not be competitive  to direct dark matter detection experiments,  though they are complementary to Borexino in the effective NGI case corresponding to region (iii). For tensor NGI, DUNE-ND has the prospect to outperform COHERENT, XENONnT and LZ for $m_T$ larger than a few MeV. Finally for the axial vector case, DUNE-ND is expected to dominate over all neutrino scattering experiments offering also complementary constraints to  existing ones from NA64 and collider measurements.

\begin{figure}[t]
\centering
\includegraphics[width = 0.49 \textwidth]{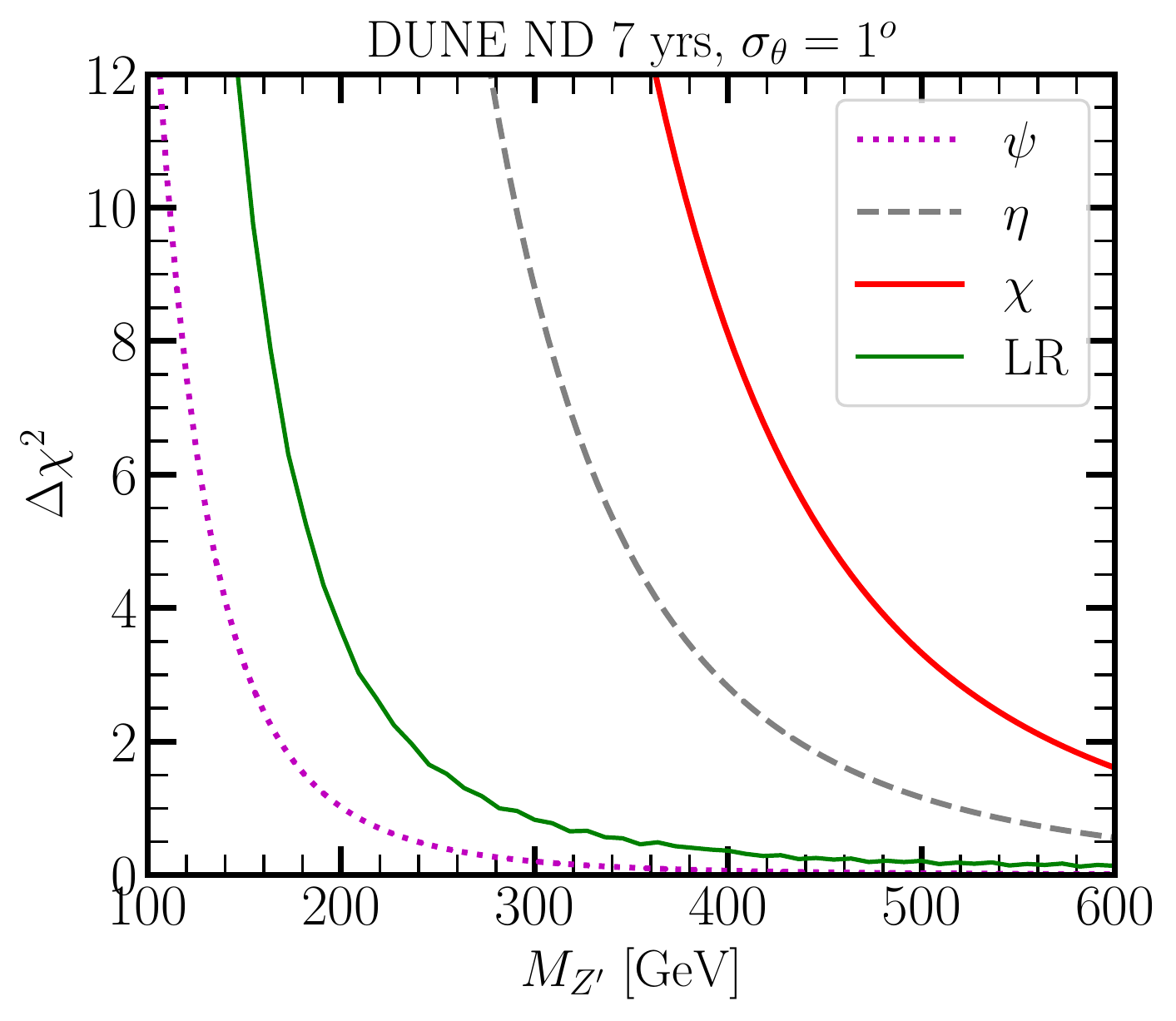}
\includegraphics[width = 0.49 \textwidth]{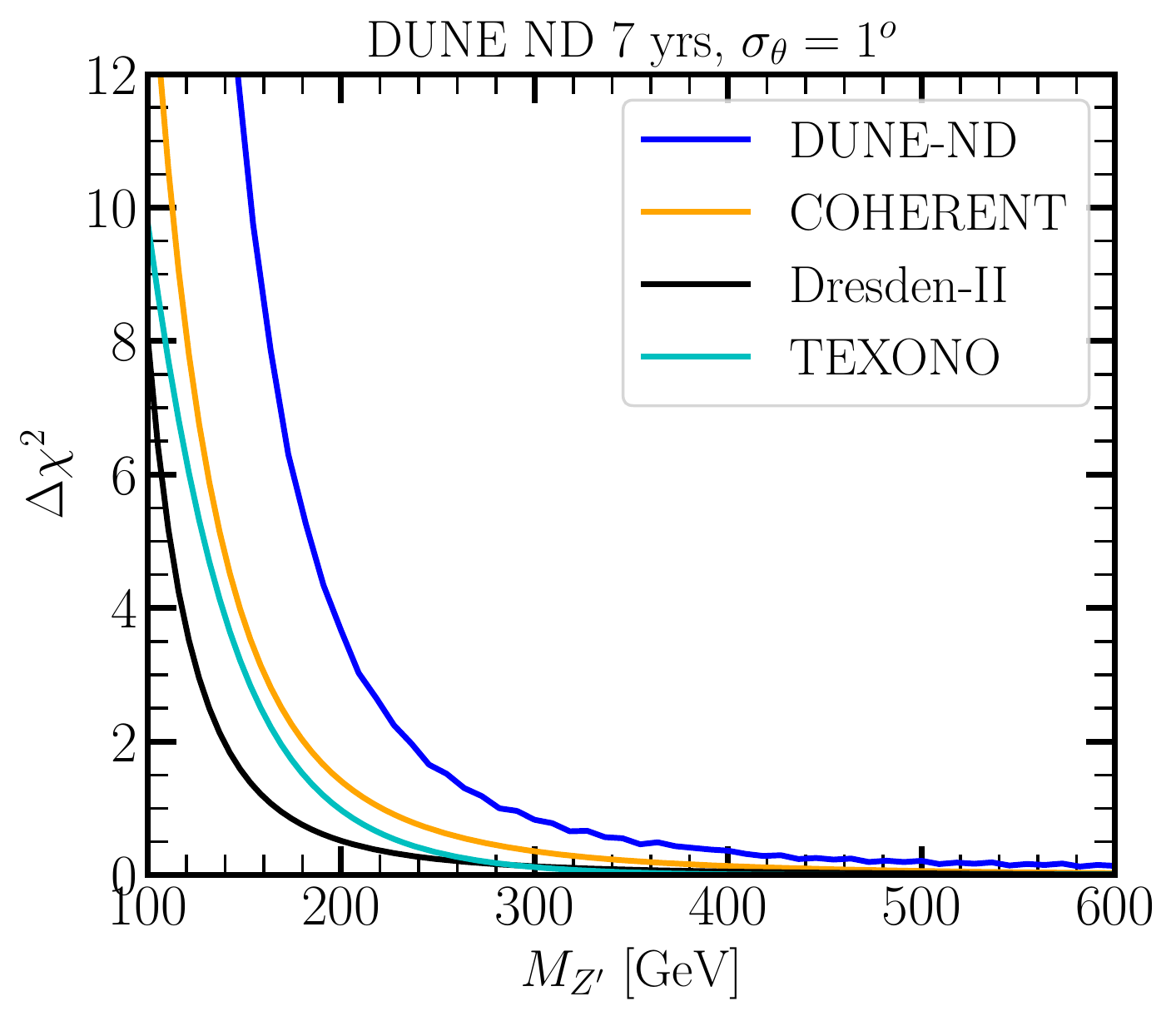}
\includegraphics[width = 0.49 \textwidth]{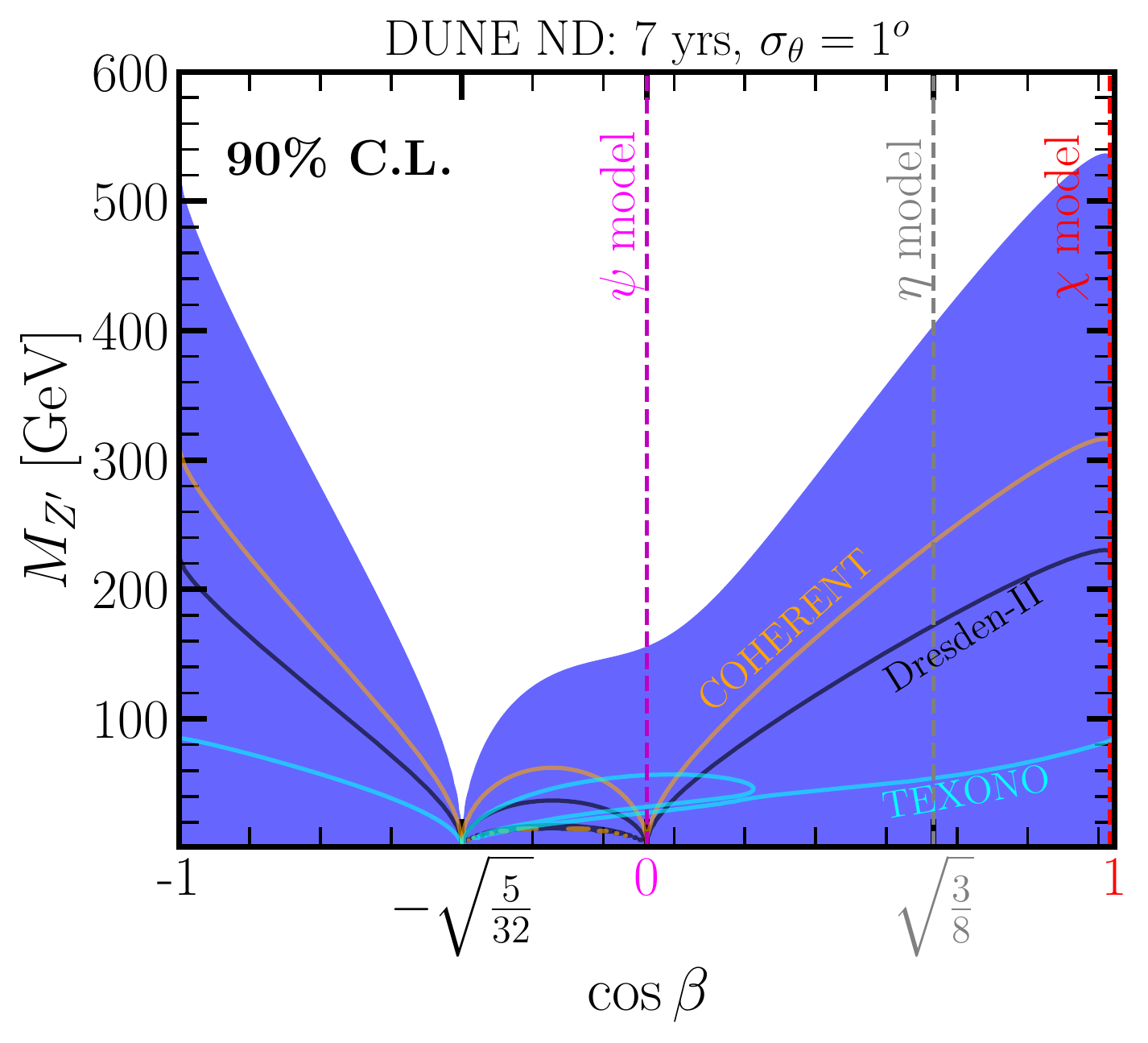}
\caption{\textbf{Upper-left:} $\Delta \chi^2$ profiles of the $Z^\prime$ mass in the LR Symmetric and $E_6$ models  assuming 3.5 yr (neutrino) + 3.5 yr (antineutrino) mode running time.  \textbf{Upper-right:} Comparison of the projected DUNE-ND sensitivities with COHERENT, Dresden-II and TEXONO.  \textbf{Lower panel:} Contours at 90\% C.L. in the $\cos \beta$-$M_{Z^\prime}$ space.}
\label{fig:LR_E6}
\end{figure}

\begin{table}[t!]
\resizebox{0.8\textwidth}{!}{%
\begin{tabular}{|l|c|cc|c|c|}
\hline
\textbf{model}  & \textbf{Spallation Source}  & \multicolumn{2}{c|}{\textbf{Reactors}}                     & \textbf{LBNF}     & \textbf{Current Limit}   \\ 
                & COHERENT (CsI+LAr) & \multicolumn{1}{c}{Dresden-II} & TEXONO & DUNE-ND & [PDG 2022]\\ \hline
$\boldsymbol{\psi}$ & --                          & \multicolumn{1}{c}{--}                  & 10              & 156 & 4560\\ \hline
$\boldsymbol{\eta}$ & 237                         & \multicolumn{1}{c}{172}                 & 54              & 404 & 3900\\ \hline
$\boldsymbol{\chi}$ & 310                         & \multicolumn{1}{c}{225}                 & 85              & 526 & 4800\\ \hline
\textbf{LR}     &      165                           & \multicolumn{1}{c}{129}                   & 155 
& 217       & 1162       \\ \hline
\end{tabular}
}
\caption{Sensitivity of DUNE-ND,  COHERENT, Dresden-II and TEXONO at 90\% C.L. on the $Z^\prime$ mass (in units of GeV) obtained in the present work. For comparison, also shown are the current limits set from collider searches at the LHC~\cite{ParticleDataGroup:2022pth}.}
\label{tab:limits}
\end{table} 

We are now turning our attention on exploring the sensitivity of DUNE-ND on LR and the various $E_6$ models. In this case, the phenomenological parameter of interest is the mediator mass entering through the parameter $\gamma= (M_{Z^0}/M_{Z^\prime})^2$. By performing a $\chi^2$ analysis in the same spirit as previously, we obtain the sensitivity profiles depicted in the left panel of Fig.~\ref{fig:LR_E6}. As can be seen the least (most) stringent constraints on the mediator mass correspond to the  $\psi$ ($\chi$) case of $E_6$ symmetry. In the right panel of Fig.~\ref{fig:LR_E6} we demonstrate a comparison of the LR sensitivities between DUNE-ND and the other neutrino scattering experiments explored in the present study such as COHERENT, Dresden-II and TEXONO. We find that DUNE-ND will be able to place the most stringent constraint on $Z^\prime$ mass.  Let us also note that for the case of \cevns the relevant expressions for calculating the $E_6$ and LR contributions are given in the Appendix~\ref{appendix2}. 

In the lower panel of Fig.~\ref{fig:LR_E6} we explore the full spectrum of $E_6$ models by allowing the value of $\cos \beta$ to vary freely. The corresponding allowed areas extracted in the present work by analyzing the COHERENT, Dresden-II and TEXONO data are superimposed for comparison. Likewise the LR case, the projected DUNE-ND sensitivities are promising to improve by a factor $\sim 2$ compared to those of the latter experiments. As can be seen there is a sharp sensitivity loss  for $\cos \beta=- \sqrt{5/32}$ due to the cancellation involved in both \eves and \cevns cross sections as explained previously. Moreover, the COHERENT and Dresden-II experiments have zero sensitivity on the $\psi$ model since another cancellation is involved in the \cevns cross section for $\cos \beta=0$ on the relevant couplings, see e.g.  Eqs.(\ref{eq:CEvNS_couplings}), (\ref{eq:E6-couplings}) in Appendix~\ref{appendix2} and Ref.~\cite{Miranda:2020zji}. Hence, we conclude that DUNE-ND is not only expected to improve over previous  neutrino-based constraints but at the same time it is clearly complementary to existing \cevns measurements. Before closing, a summary of our results regarding the DUNE-ND sensitivity on $E_6$ and LR models along with a comparison with the rest neutrino experiments analyzed in the present work is given in Table~\ref{tab:limits}.

Before closing we would like to emphasize the complementarity of our present results with existing ones probed by collider searches at the LHC~\cite{ParticleDataGroup:2022pth}. 
Specifically, we show how neutrino data coming from \eves and \cevns measurements can be utilized to explore the $Z^\prime$ mass as an alternative research channel to the dilepton $Z^\prime$ decay explored at the LHC. 
More importantly, we stress that collider searches are relevant in the high energy window e.g. Refs.~\cite{ATLAS:2019erb,CMS:2021ctt} provide limits for $M_{Z^\prime} > 250$~GeV, Ref.~\cite{Electroweak:2003ram} for $M_{Z^\prime} > 200$~GeV, and Ref.~\cite{Kang:2004bz} for $M_{Z^\prime} > 100$~GeV, thus leaving the low-energy region unexplored which is what our present analysis using neutrino data focuses on.

\section{Conclusions}
\label{sec:conclusions}

In this work we have simulated E$\nu$ES-induced signals in the context of model-independent NGIs that are expected to trigger the DUNE-ND. To this purpose, we performed Monte Carlo simulations of the reconstructed signal in $E_e \theta_e^2$ space taking into account detector-specific effects such as angular resolution, systematic uncertainties and realistic backgrounds on the analysis of On-Axis and Off-Axis spectra. For the first time, we have examined the prospects of constraining light axial-vector, scalar, pseudoscalar and tensor mediators at the ND complex using \eves events. Moreover, we considered particular models such as those arising from $U(1)_{B-L}$,  $U(1)_{L_\mu - L_\tau}$, LR  and $E_6$ gauge symmetries. We have furthermore performed a comparison of the projected DUNE-ND sensitivities with additional experimental probes. In particular we considered the \eves data from the reactor neutrino experiment TEXONO, \cevns data from COHERENT and Dresden-II, solar neutrino data from Borexino, low-energy data from direct dark matter detection experiments such as XENONnT and LZ as well as high-energy collider data. In all cases we find that by exploiting the intense LBNF neutrino beam in conjunction with the multi-ton Liquid Argon detector, DUNE-ND has promising prospects to constrain a considerable part of the parameter space in question, offering complementary results to low-energy neutrino and direct dark matter detection experiments. We demonstrated that future DUNE-ND data will offer a powerful tool for placing the most stringent constraints in the range $m_X >100~\mathrm{MeV}$ up to few GeV  regarding NGIs.   Finally, regarding LR and $E_6$ symmetric models, we showed that DUNE and COHERENT are drastically improving previous limits on the vector boson mass $M_{Z^\prime}$ set by TEXONO in the low-energy regime, and we also highlighted their complementarity to collider searches.

\section*{Acknowledgements} 
We are indebted to Yuber F. Perez-Gonzalez for valuable discussions regarding the event rate reconstruction procedure. We also thank the authors of Ref.~\cite{Chakraborty:2021apc} for useful correspondence. DKP thanks O. Miranda and A. Khan for their insightful comments on the manuscript.
This work was supported by the Hellenic Foundation for Research and Innovation (H.F.R.I.) under the “3rd Call for H.F.R.I. Research Projects to support Post-Doctoral Researchers” (Project Number: 7036).

\noindent

\appendix
\section{\label{appendix1}Normalized PDF event spectra at DUNE-ND}

\begin{figure}[ht]
\centering
\includegraphics[width = 0.32 \textwidth]{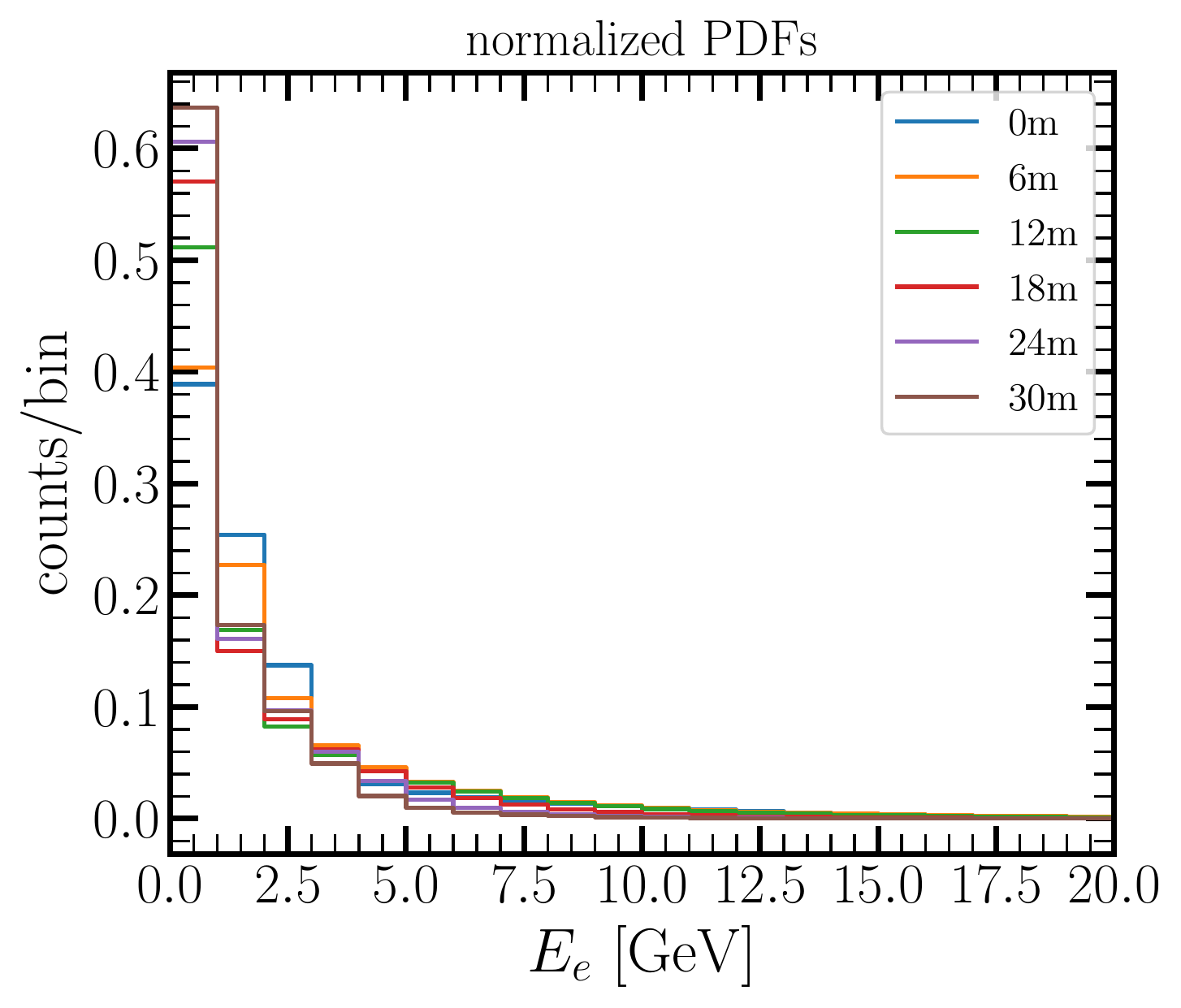}
\includegraphics[width = 0.32 \textwidth]{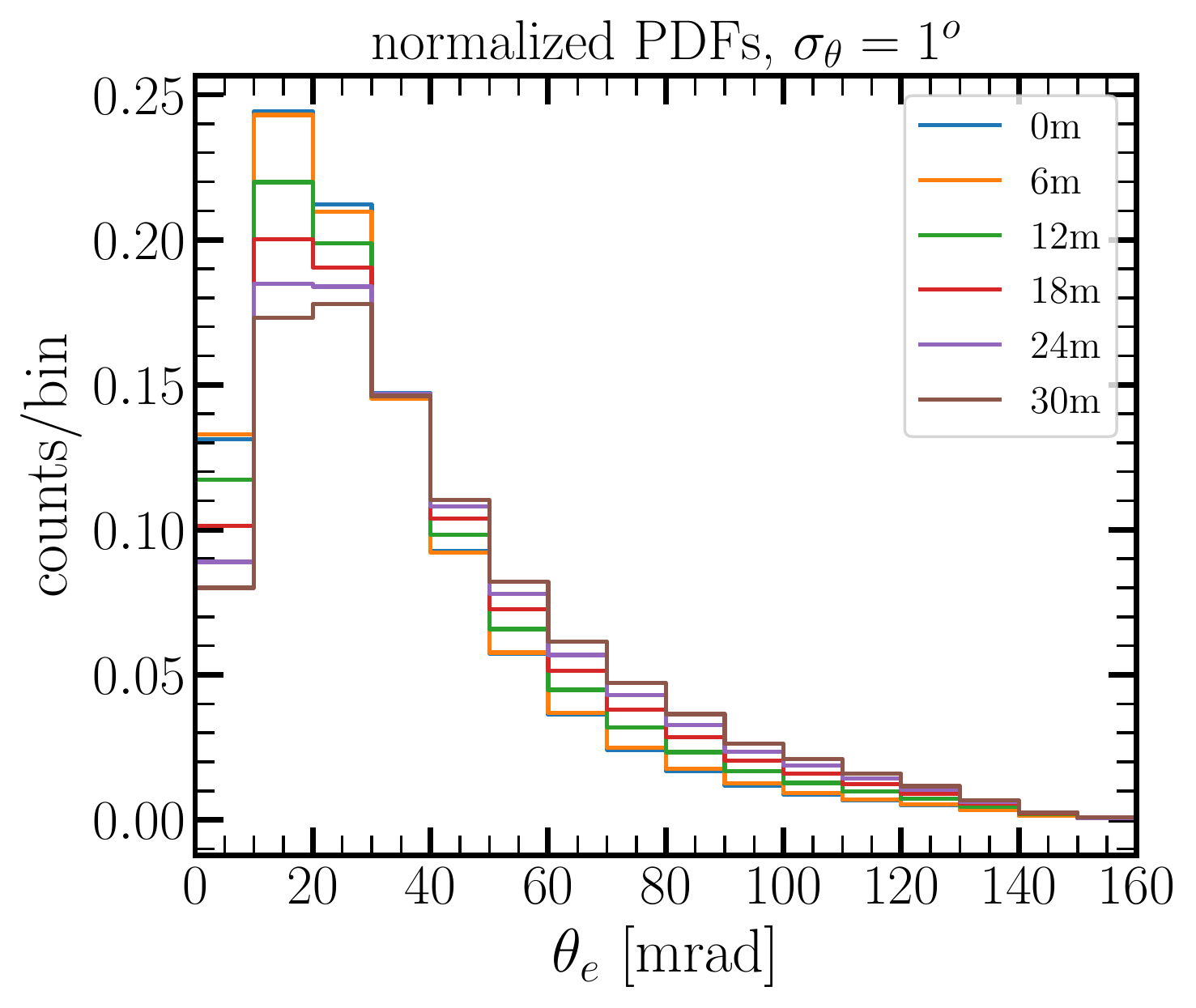}
\includegraphics[width = 0.32 \textwidth]{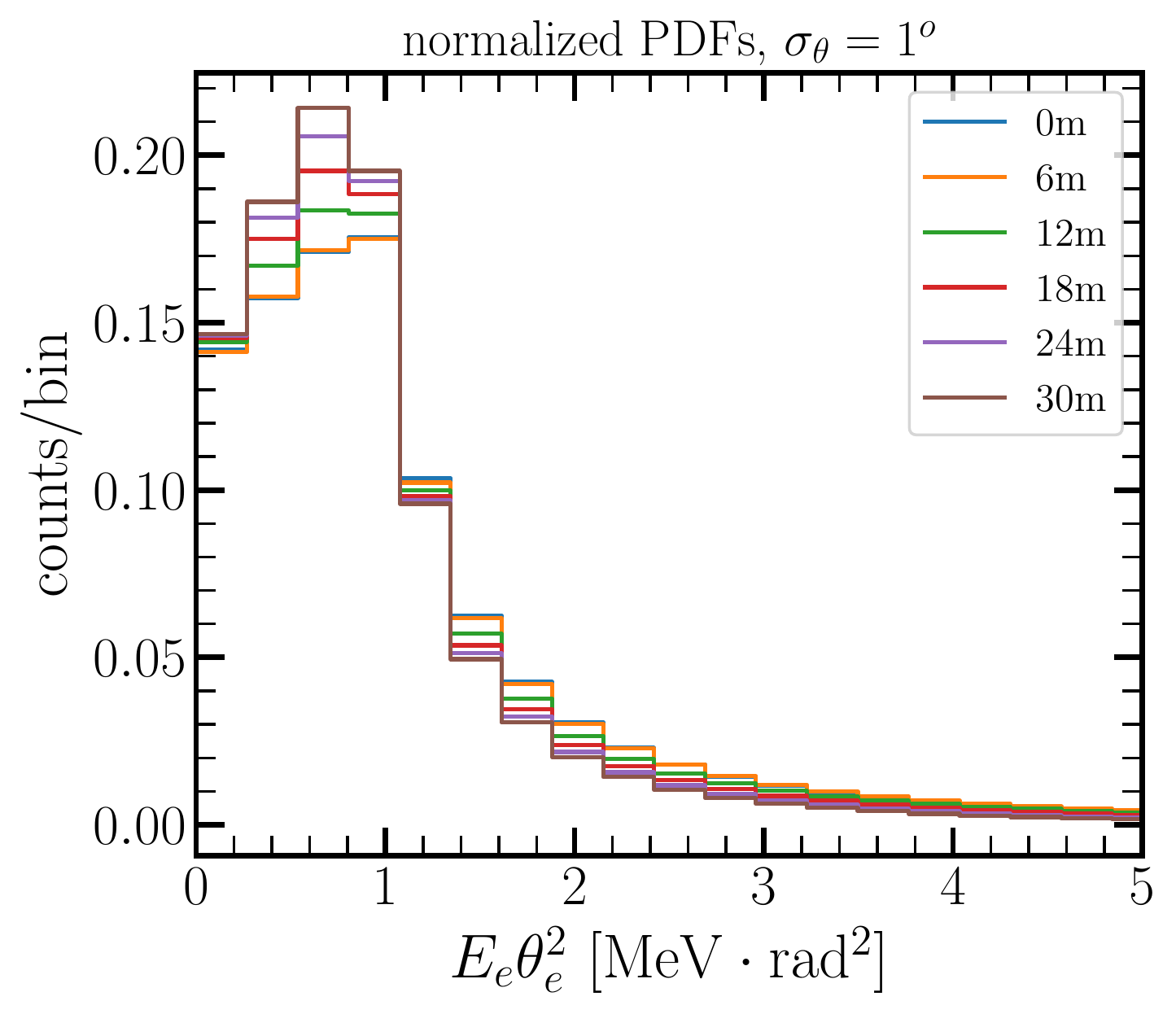}
\caption{Normalized PDFs for the event spectra in the SM assuming  an exposure of 3.5 yr  in neutrino mode at the DUNE-ND. The results are projected in electron energy $E_e$ (left), scattering angle $\theta_e$ (center) and  $E_e \theta_e^2$ space (right).}
\label{fig:SMeventsPDF}
\end{figure}

Figure~\ref{fig:SMeventsPDF} shows the reconstructed SM \eves signal events as a function of the electron energy $E_e$ (left), scattering angle $\theta_e$ (center) and  $E_e \theta^2$ space (right), after 3.5 years of data collection in neutrino mode. The results are presented for the various On-Axis and Off-Axis locations, as done in Figs.~\ref{fig:SMevents} and \ref{fig:SMevents_recoil_theta} . However, here they are illustrated as  normalized PDFs in order to appreciate the spectral shape differences among the various detector locations. As expected, when projecting to $E_e$, the On-Axis (30 m) event spectra appear as having the wider (narrower) distribution since the On-Axis neutrino energy distribution peaks at higher energy. The opposite behavior is found in the scattering angle projected PDF spectra. The latter is  understood since the On-Axis event spectra are the most boosted ones, and therefore they will be detected with smaller scattering angles. Finally the $E_e \theta_e^2$ projection appears as the combination of the $E_e$ and $\theta_e$ distributions.

\section{\label{appendix2}Relevant expressions for CE$\nu$NS}

The SM \cevns cross section in terms of the nuclear recoil energy $T_A$ reads
\begin{equation}
\left(\frac{d \sigma}{dT_A}\right)_{\text{SM}} = \frac{G_F^2 M}{\pi}  \mathcal{Q}_V^2 \left(1 - \frac{M T_A}{2 E_\nu^2}\right)  \, ,
\label{eq:xsec-cevns}
\end{equation}
with $M$ being the nuclear mass, $|\vec{\mathfrak{q}}|= \sqrt{2 M T_A}$ denotes the magnitude of the three-momentum transfer and 
$\mathcal{Q}_V$ is the vector weak charge written as
\begin{equation}
\mathcal{Q}_V =  \left[ 2(g_{u}^{L} + g_{u}^{R}) + (g_{d}^{L} + g_{d}^{R}) \right] Z F_Z(|\vec{\mathfrak{q}}|^2)  + \left[ (g_{u}^{L} + g_{u}^{R}) +2(g_{d}^{L} + g_{d}^{R}) \right] N  F_N(|\vec{\mathfrak{q}}|^2) \, .
\label{eq:CEvNS_couplings}
\end{equation}
Here, $F_Z(|\vec{\mathfrak{q}}|^2)$ and $F_N(|\vec{\mathfrak{q}}|^2)$ are the nuclear form factors for protons and neutrons, respectively, while the vector weak charge $\mathcal{Q}_V$ is expressed in terms of the $P$-handed couplings  for the quark $q=\{u,d\}$  as
\begin{equation}
\begin{aligned}
g_{u}^{L} =&  \left( \frac{1}{2}-\frac{2}{3} \sin^2 \theta_W \right)  \, ,\\
g_{d}^{L} =&  \left( -\frac{1}{2}+\frac{1}{3} \sin^2 \theta_W \right) \, ,\\
g_{u}^{R} =&  \left(-\frac{2}{3}  \sin^2 \theta_W \right)   \, ,\\
g_{d}^{R} =&  \left(\frac{1}{3}  \sin^2 \theta_W \right) \, . 
\end{aligned}
\end{equation}
In the context of LR symmetry, the SM couplings $g_q^P$ are substituted by the corresponding $f_q^P$ as~\cite{Polak:1991pc}
\begin{equation}
\begin{aligned}
f^{L}_u=&  \mathcal{A} g_{u}^{L} + \mathcal{B} g_{u}^{R} \, , \\ 
f^{L}_d=&\mathcal{A} g_{d}^{L} + \mathcal{B} g_{d}^{R} \, ,\\
f^{R}_u=&  \mathcal{A} g_{u}^{R} + \mathcal{B} g_{u}^{L} \, , \\ 
f^{R}_d=&\mathcal{A} g_{d}^{R} + \mathcal{B} g_{d}^{L} \, ,\\
\end{aligned}
\end{equation}
where the $\mathcal{A}$ and $\mathcal{B}$ parameters are defined as in Eq.(\ref{eq:LR_params}).

For the case of the $E_6$ model, the SM \cevns couplings are modified as follows
\begin{equation}
\begin{aligned}
f_{u}^{L} =& g_{u}^{L} + \varepsilon_u^L \, ,\\
f_{d}^{L} =& g_{d}^{L} + \varepsilon_d^L \, ,\\
f_{u}^{R} =& g_{u}^{R} + \varepsilon_u^R \, ,\\
f_{d}^{R} =& g_{d}^{R} + \varepsilon_d^R \, , 
\end{aligned}
\end{equation}
where the new physics is encoded in the $\varepsilon_{q}^{P}$ couplings according to~\cite{Barranco:2007tz}
\begin{equation}
\begin{aligned}
\varepsilon_u^L=&-4 \gamma \sin^2 \theta_W \left(\frac{c_{\beta}}{\sqrt{24}}-\frac{s_{\beta}}{3} \sqrt{\frac{5}{8}}\right)\left(\frac{3 c_{\beta}}{2 \sqrt{24}}+\frac{s_{\beta}}{6} \sqrt{\frac{5}{8}}\right) \, , \\
\varepsilon_d^R=&-8 \gamma \sin^2 \theta_W \left(\frac{3 c_{\beta}}{2 \sqrt{24}}+\frac{s_{\beta}}{6} \sqrt{\frac{5}{8}}\right)^{2} \, , 
\\
\varepsilon_d^L=& \varepsilon_u^L=-\varepsilon_u^R \, .
\end{aligned}
\label{eq:E6-couplings}
\end{equation}
%

\bibliographystyle{utphys}
\bibliography{bibliography}

\end{document}